\documentclass[10pt]{article}

\usepackage{amsmath}
\usepackage{amssymb}

\usepackage{graphicx}

\usepackage{cite}

\usepackage{color} 

\usepackage[labelfont=bf,labelsep=period,justification=raggedright]{caption}

\makeatletter
\renewcommand{\@biblabel}[1]{\quad#1.}
\makeatother

\date{}

\begin{document}

\begin{flushleft}
{\Large
\textbf{Web search queries can predict stock market volumes.}
}
\\
Ilaria Bordino$^{1}$, 
Stefano Battiston$^{2}$, 
Guido Caldarelli$^{3,4,5}$,
Matthieu Cristelli$^{3,\ast}$,
Antti Ukkonen$^{1}$,
Ingmar Weber$^{1}$
\\
\bf{1} Yahoo! Research, Avinguda Diagonal 177, Barcelona, Spain
\\
\bf{2} ETH Chair of System Design, Kreutzplatz 5, Zurich Switzerland
\\
\bf{3} Inst. of Complex Systems CNR, Dip. Fisica, ``Sapienza'' Univ., P.le Moro 5 00185 Rome, Italy
\\
\bf{4} London Institute for Mathematical Sciences, South Street 22, Mayfair London, UK
\\
\bf{5} IMT, Institute for Advanced Studies, Piazza S. Ponziano, 6, 55100 Lucca, Italy
\\
$\ast$ E-mail: matthieu.cristelli@roma1.infn.it
\end{flushleft}

\section*{Abstract}
We live in a computerized and networked society where many of our
  actions leave a digital trace and affect other people's
  actions. This has lead to the emergence of a new data-driven
  research field: mathematical methods of
  computer science, statistical physics and sociometry provide
  insights on a wide range of disciplines
  ranging from social science to human
  mobility.  A recent important discovery
  is that search engine traffic
  (i.e., the number of requests submitted by
  users to search engines on the www) can be used to track and, in
  some cases, to anticipate the dynamics of social
  phenomena. Successful examples include unemployment
  levels, car and home sales, and epidemics spreading.
  Few recent works applied this approach to stock
  prices and market
  sentiment.
  However, it remains unclear if trends in financial markets can be
  anticipated by the collective wisdom of on-line users on the web.
  Here we show that daily trading volumes of stocks traded in NASDAQ-100 are
  correlated with daily volumes of queries related to the same stocks.
  In particular, query volumes anticipate in many cases peaks of
  trading by one day or more. Our analysis is carried out on a unique
  dataset of queries, submitted to an important web search engine,
  which enable us to investigate also the user behavior. We show that
  the query volume dynamics emerges from the collective but seemingly
  uncoordinated activity of many users.
  These findings contribute to the debate on the identification of
  early warnings of financial systemic risk, based on the activity of
  users of the www.
\section*{Author Summary}

\section*{Introduction}
Nowadays many of our activities leave a digital trace: credit card 
transactions, web activities, e-commerce, mobile-phones, GPS navigators, etc.
This networked reality has favored the emergence of a new data-driven
 research field 
  where mathematical methods of
  computer science \cite{mitchell}, statistical physics \cite{vesp1} and sociometry 
  provide effective insights on a wide range of disciplines like \cite{evans2010machine}
  social sciences \cite{lazer2009life}, human mobility \cite{gonzalez2008mobility}, etc.
  
\begin{figure}[!ht]
\begin{center}
\includegraphics[scale=0.62]{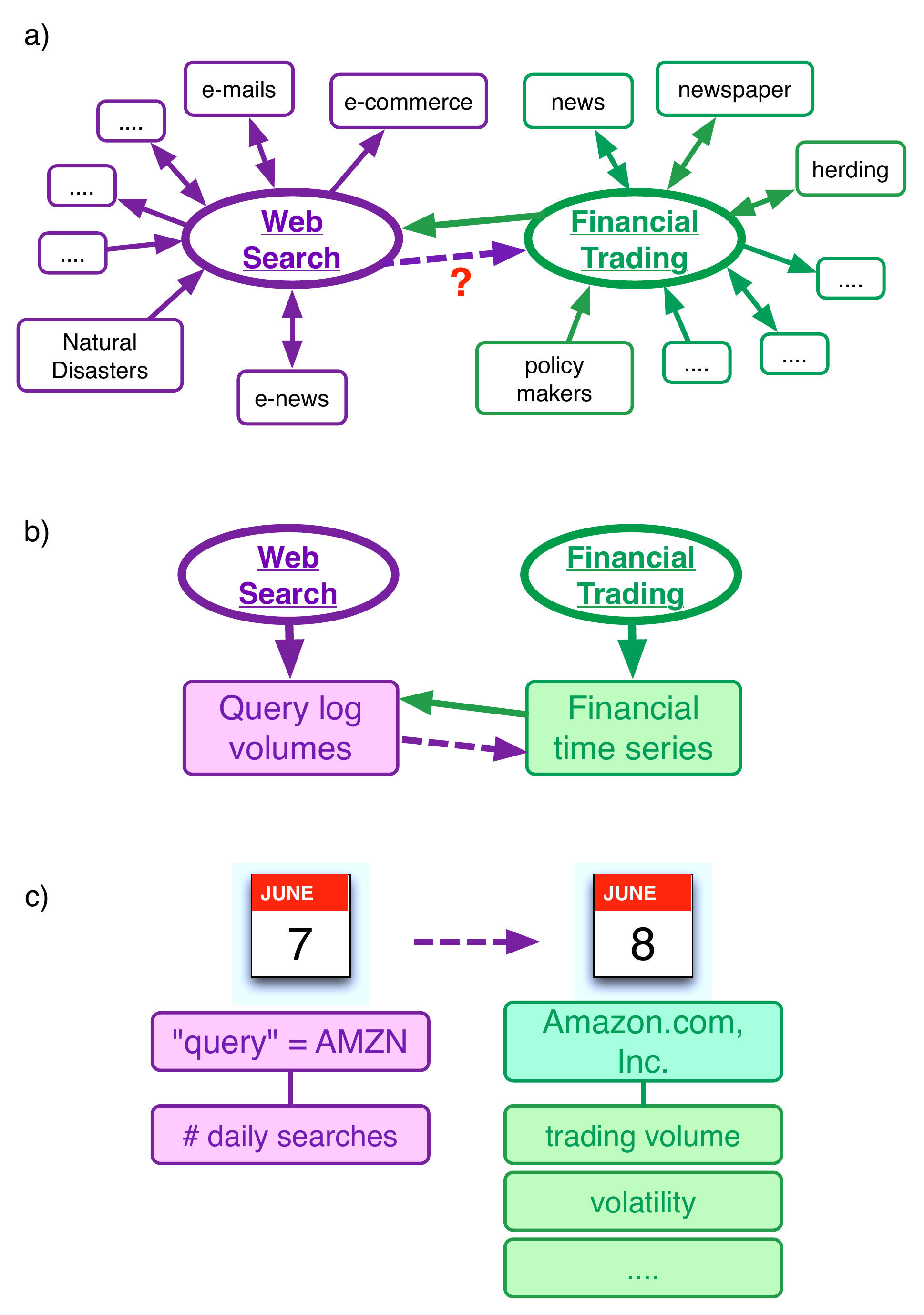}
\end{center}
\caption{\textbf{Graphical illustration of the analysis presented in this paper.} 
The study of queries is gaining more and more attention as an important tool for the understanding of social and financial systems. Users perform web searches in order to collect news or browse e-newspaper sites. In particular local or global events such as natural disasters can generate local or global waves of searches through the web. As a result, the logs of these search-engines' queries are an unprecedented source of anonymized information about human activities. 
In this paper we provide a detailed analysis on a particular application of these ideas; that is,  
the anticipation of market activity from user queries. This picture graphically summarizes our procedure. 
In particular, we investigate which is the relationship between web searches and market movements and whether web searches \textit{anticipate} market activity.  While we can expect that large fluctuations in markets, produce spreading of news or rumors or government's actions and therefore induce web searches  (solid green arrow in panel $a$), 
we would like to check if web searches affect or even anticipate financial activity (broken violet arrow in panel $a$). 
In detail we investigate if today's query volumes about financial stocks somehow anticipate financial indicators of tomorrow such as trading volumes, daily returns, volatility, etc, (panels $b$ and $c$) and we find a significant anticipation for trading volumes.}
\label{figure0}
\end{figure}

Recent investigations showed that Web search traffic can be used to
accurately track several social phenomena \cite{choi2009trends,goel2010consumer,twitter,mob}.
One of the most successful results in this direction, concerns the
epidemic spreading of influenza virus among people in the USA. It has
been shown that the activity of people querying search engines for
keywords related to influenza and its treatment allows to anticipate
the actual spreading as measured by official data on contagion
collected by Health Care Agencies \cite{ginzberg2009epidemics}.
In this paper, we address the issue whether a similar approach can be
applied to obtain early indications of movements in the financial
markets \cite{saavedra2011traders,preis2010query,bollen2011twitter} (see Fig. \ref{figure0} for a graphical representation of this issue). 
Indeed, financial turnovers, financial contagion and, ultimately,
crises, are often originated by collective phenomena such as herding
among investors (or, in extreme cases, panic) which signal the
intrinsic complexity of the financial system \cite{bouchaud2009economy}.
Therefore, the possibility to anticipate anomalous collective
behavior of investors is of great interest to policy makers \cite{HM11,SFSVVW09,bouchaudnat} because
it may allow for a more prompt intervention, when this is appropriate. 
For instance the authors of \cite{suggest1} predict economical outcomes starting from social data, however,
these predictions are not in the context of financial markets.\\
Furthermore it has been shown how volume shifts can be correlated with 
price movements \cite{stanley2,stanley3,stanley1}.

Here, we focus on queries submitted to the Yahoo! search
engine that are related to companies listed on the NASDAQ stock exchange.
Our analysis is twofold. On the one hand, we assess the relation over
time between the daily number of queries (``query volume'', hereafter)
related to a particular stock and the amount of daily exchanges over the
same stock (``trading volume'' hereafter). We do so by means not only
of a time-lagged cross-correlation analysis, but also by means of the
Granger-causality test.  
On the other hand, our unique data set allows us to analyze the
search activity of individual users in order to provide insights
into the emergence of their collective behavior.

\section*{Results}

\begin{figure}[!ht]
\begin{center}
\includegraphics[scale=0.4]{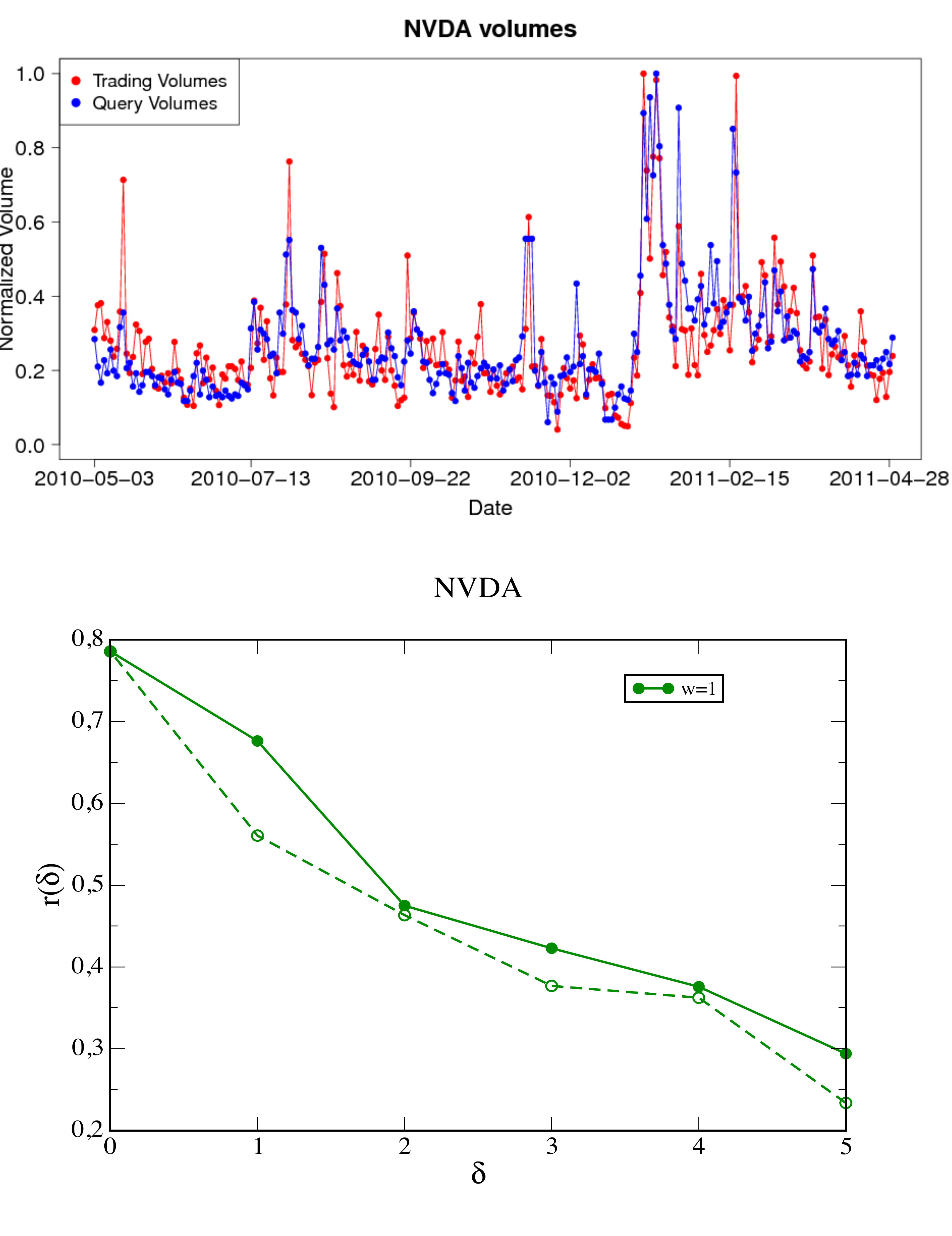}
\end{center}
\caption{\textbf{Query log volumes and trading volumes: cross correlation analysis (ticker: ``NVDA'')}. \textbf{(up)} 
Time evolution of normalized query-logs volumes for the ticker ``NVDA'' compared with the 
trading-volume of the ``NVIDIA Corporation''. The data for both query-logs (blue) and trading volume (red) 
are aggregated on a daily basis. \textbf{(bottom)} The plot of the sample cross correlation function 
$r(\delta)$ as defined in Eq. (\ref{equat1}) $vs$ 
absolute values of the time lag $\delta$ (positive values of $\delta$ correspond to solid lines 
while negative values of the time lag correspond to the broken lines). The correlation coefficients 
at positive time lags are always larger than the corresponding at negative ones, this suggests that 
today's query volumes anticipate and affect the trading activity of the following days 
(typically one or two days at most).}
\label{figure1}
\end{figure}

\begin{figure}[!ht]
\begin{center}
\includegraphics[scale=0.4]{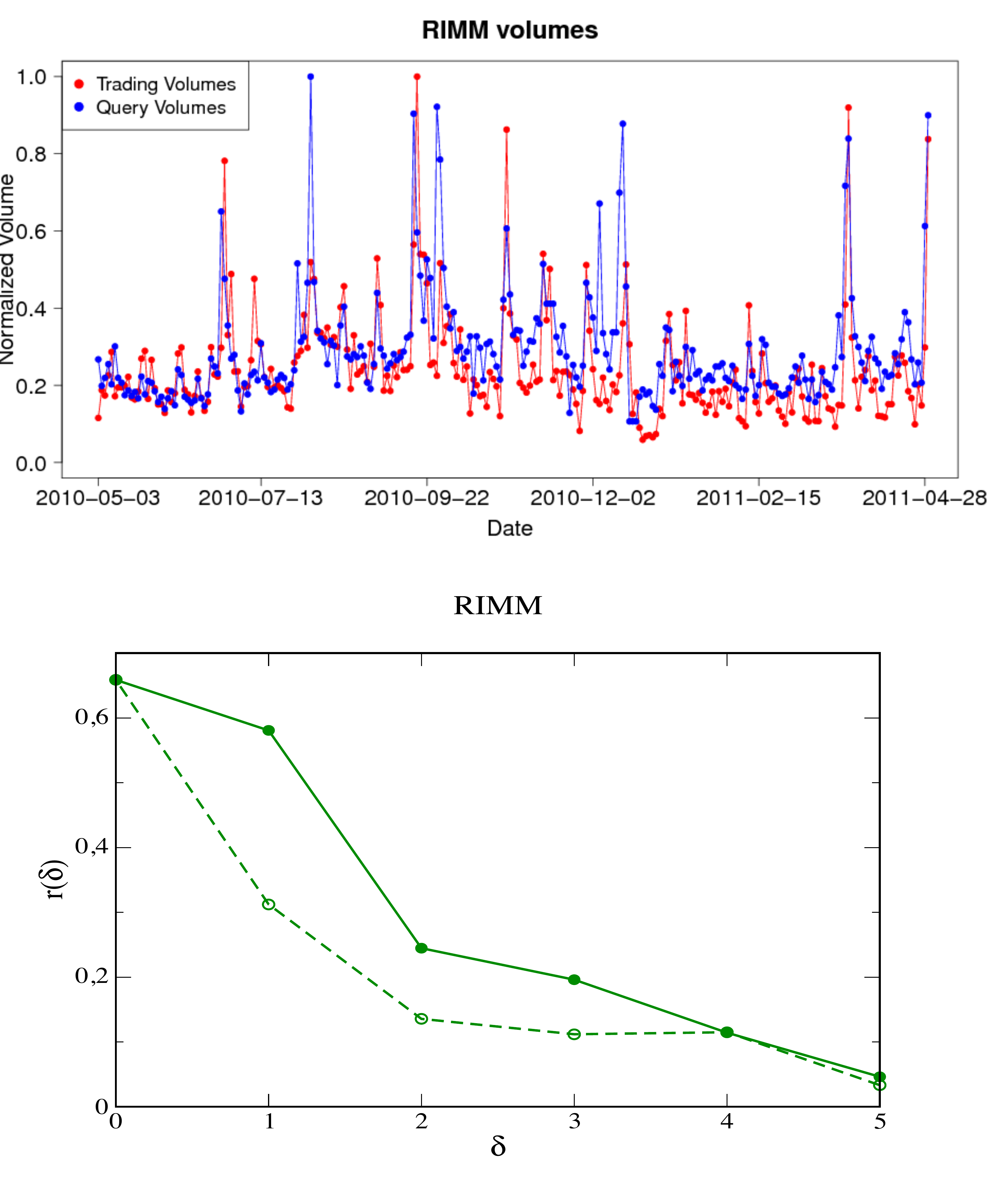}
\end{center}
\caption{\textbf{Query log volumes and trading volumes: cross correlation analysis  (ticker: ``RIMM'')}. \textbf{(up)} 
Time evolution of normalized  query-logs volumes for the ticker ``RIMM'' compared with the 
trading-volume of the ``Research In Motion Limited''. The data for both query-logs (blue) and trading volume (red) 
are aggregated on a daily basis. \textbf{(bottom)} The plot of the sample cross correlation function 
$r(\delta)$ as defined in Eq. (\ref{equat1}) vs 
absolute values of the time lag $\delta$ (positive values of $\delta$ correspond to solid lines 
while negative values of the time lag correspond to the broken lines). As in the case of the ticker ``NVDA''
corresponding to the company ``NVIDIA Corporation'' in Fig. \ref{figure1}, the correlation coefficients 
at positive time lags are always larger than the corresponding at negative ones, this suggests that 
today's query volumes anticipate and affect the trading activity of the following days 
(typically one or two days at most).}
\label{figure2}
\end{figure}

\begin{figure}[!ht]
\begin{center}
\includegraphics[scale=0.3,angle=90]{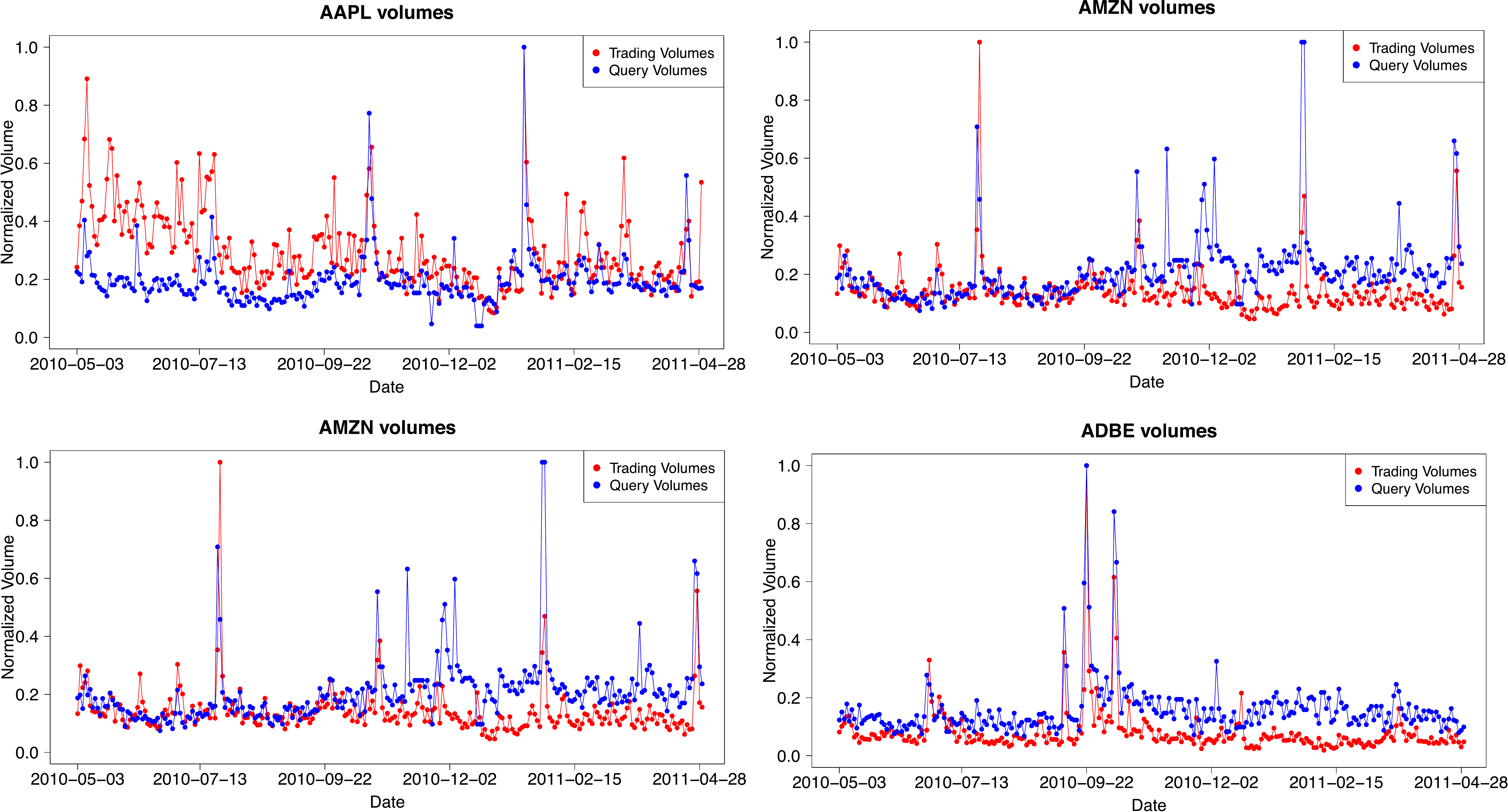}
\end{center}
\caption{\textbf{Query volumes and trading volumes}. We plot the query-search volumes and trading volumes time series
for four stocks (AAPL, AMZN, NFLX and ADBE) to show that the patterns observed in Figs. \ref{figure1} and \ref{figure2} are common to most of stocks of the set considered (NASDAQ-100).}
\label{fig:volumes}
\end{figure}

In our analysis we consider a set of companies (``NASDAQ-100 set''
hereafter) that consists of the companies included in the NASDAQ-100
stock market index (the 100 largest non-financial companies traded on
NASDAQ). We list these companies  in Table \ref{table:1}. 
Previous studies \cite{preis2010query} looked at
stock prices at a weekly time resolution and found that the volume of
queries is correlated with the volume of transactions for all stocks
in the S\&P 500 set for a time lag of $\Delta t = 0$ week, i.e. the
present week query volumes of companies in the S\&P 500 are
significantly correlated with present week trading volumes of the S\&P
500\footnote{in addition, differently from \cite{preis2010query} we
  use daily data from Yahoo! search engine and we look at query
  volumes from single stocks and do not aggregate these volumes}. The
authors of \cite{preis2010query} suggest that the query volume can be interpreted as reflecting the attractiveness of trading a stock. Further, they find that this attractiveness effect lasts for several weeks and, citing the authors of \cite{preis2010query}, \textit{present price movements seem to influence the search volume in the following weeks} pointing out that new analysis on data at a smaller time scale are needed.\\ 
This last observation is the starting point of the present work. Is it
possible to better investigate the relation between search traffic and
market activity on a daily time scale? And, even more important, can
query volumes anticipate market movements and be a proxy for market
activity? In other words in this paper we are addressing the question
whether web searches can be a forecasting tool for financial markets
and not only a nowcasting one. This is a novel analysis which try to
quantify the link and the direction of the link between search traffic and financial activity.\\
We consider search traffic as well as market activity at a daily frequency and find a strong
correlation between query volumes and trading volumes for all stocks
in the NASDAQ-100 set. 
Fig. \ref{figure1} (top panel) shows the time evolution of the query volume of the
ticker ``NVDA'' and the trading volume of the corresponding company
stock ``NVIDIA Corporation'' and Fig. \ref{figure2} (top panel) shows the same
plot for query volume of the ticker ``RIMM'' and the trading volume of
the company stock ``Research In Motion Limited'' (see also Section
``Materials and Methods''). A simple visual inspection of these figures (see also Fig. \ref{fig:volumes}) reveals a
clear correlation between the two time series because peaks in one
time series tend to occur close to peaks in the other.\\
%
The lower panels of Figs. \ref{figure1} and \ref{figure2} report the values of cross correlation
between trading and query volume as a function of the time lag
$\delta$ defined as the time-lagged Pearson cross correlation $r(\delta)$ coefficient
between two time series $Q_t$ and $T_t$:
\begin{equation}\label{equat1}
r(\delta)=\frac{\sum_{t=1}^{n}(Q_t-\overline{Q})(T_{t+\delta}-
\overline{T})}{\sqrt{\sum_{t=1}^{n}(Q_t-\overline{Q})^2}\sqrt{\sum_{t=1}^{n}(T_{t+\delta}-
\overline{T})^2}}
\end{equation}
where $\overline{Q}$, $\overline{T}$ are the sample averages of the two
time series (in this case $Q$ and $T$ represent query and trading
volumes, respectively). The coefficient $r(\delta)$ can range from $-1$ (anticorrelation) to $1$ (correlation).

The cross correlation coefficients for positive values of $\delta$
(solid lines) are always larger than the ones for negative time lag
(broken lines). This means that query volumes tend to anticipate
trading volumes. Such an anticipation spans from $1$ to $3$ days at most. \\
Beyond a lag of $3$ days, the correlation of query volumes
with trading volumes vanishes. In Table \ref{tbl:clean-ticker} where we report the cross correlation function between queries and trading volumes averaged over the 87 companies in the NASDAQ-100 for which we have a clean query-log signal. In Table \ref{tbl:best} instead we report the cross correlation functions for some of the 87 companies investigated in Table \ref{tbl:clean-ticker} (for the sake of completeness in the Supporting Information in Tables S1 and S2 we report the cross correlation functions for all the clean stocks while in Table S3  the cross correlation functions for those stocks characterized by spurious origin of the query volume). 

As a first result from this analysis we find that the significant correlation between query volumes and trading volumes at $\delta= 0$ confirms the results of \cite{preis2010query} also at a daily timescale. Our findings (i.e. positive correlation for negative time lags) also support the vision that present market activity influences future users' activity but in contrast with \cite{preis2010query} the length of this \textit{influence} appears to be much shorter than what expected (only few days). It appears that the correlation only emerges at a daily scale and seems to be not observed at weekly resolution. \\
However, the most striking result is that the cross-correlation
coefficients between present query volumes and future trading volumes appears to be larger than the coefficient of the opposite case. In the following of this paper we discuss in detail this anticipation effect and give a statistical validation of our finding.   \\

\subsection*{Statistical validation}
In order to assess the statistical significance of the results for the
NASDAQ-100 set, we construct a reshuffled data set in which the
query volume time series of a company $C_i$ is randomly paired to the
trading volume time series of another company $C_j$. The values of the
cross-correlation coefficient averaged over $1000$ permutations
(values which span the range $[-0.033,0.06]$ ) are smaller than the
original one (which is $0.31$) by a factor $10$. The residual
correlation present in the reshuffled dataset can be explained in
terms of general trends of the market and of the specific
(technological) sector considered \cite{kertesz1,kertesz2, garlaschelli2005scale-free}. 

As a second test we remove the top five (and ten) largest events from the trading volume times series 
in order to verify if the results shown in Table  \ref{tbl:best}  (the results for all the stocks are reported in Tables S4 and S5 of Supporting Information) are dominated by these events. 
In Table \ref{tbl:drop}  we report the comparison between the values of the cross correlation coefficient $r(0)$ of the two series for a selection of stocks. A significant correlation is still observed for most of the stocks considered. This important test supports the robustness of our findings. In fact, even if the drop indicates that the distributions underlying the investigated series are fat-tailed (see  Figs. S1-S6 of Supporting Information and the discussion about the validity of the Granger test in the following of the paper) and that a significant fraction of the correlation is driven by largest events (about $5\%$ of the events are responsible for $25-30\%$ of the correlation on the average), more than half of the correlation (for some stocks this percentage reaches $90\%$) cannot be explained by these extreme events.

Turning now the discussion towards the validation of the fact that query volumes anticipate trading volumes, as a first issue, it is a well-known fact that trading volumes and volatility are correlated and this last appears to be autocorrelated \cite{cont,stanl,boubook} (the decay of the volatility is well-described by a power law with an exponent ranging between $-1$ and $0$). Therefore the correlation between the query volumes and the future trading volumes shown in Figs. \ref{figure1} and \ref{figure2} could be explained in terms of these two effects. In this respect we compare the lagged cross-correlation function between a proxy for the volatility (the absolute value of price returns) and the query volumes with the results shown in Table \ref{tbl:clean-ticker}. As shown in Fig. \ref{figure3}, the $\delta>0$ branch in the volatility case is equal or even smaller than the value observed in the $\delta<0$ one, differently from the trading volume case. If the origin of the effect were due to the autocorrelation component of the volatility, we would expect a similar behavior for both cross-correlation functions. In addition we observe that the volatility autocorrelation function decays much slower (from weeks to months) than the typical time decay of the cross correlations here investigated (few days). This supports the non-autocorrelated origin of the anticipation effect.

As a second measure of the anticipation effect, we also performed a
Granger causality test \cite{granger1969causal} in order to determine if todays
search traffic provides significant information on forecasting
trading volumes of tomorrow. We find that trading volumes can be
considered Granger-caused by the query volume. We want to point out that Granger-causality does not imply 
a causality relation between the two series. In fact it can be argued
with a simple counterexample that two Granger-caused series may be
driven by a third process and therefore the interpretation of the
Granger relation as a causality link would be wrong. In our analysis
the results of the Granger test are only used to assess the direction
of the anticipation between queries and trading activity. In this
sense we claim that query volumes observed today are informative of (and consequently forecast) tomorrows trading volumes.\\
Furthermore, the fat-tailed nature of the distributions under
investigation (see Figs. S1-S6 of Supporting Information) may weaken the results of
the Granger-test which, in principle, requires gaussian distributions
for the error term of the regressions
\cite{granger1969causal}. However, we perform a series of additional analyses and tests which support and confirm the picture coming from Granger-test results (see Section ``Materials and Methods''
for further details).
\subsection*{Users' behavior}
In the second part of our investigation we focus on the activity of
single users.  We are able to track the users who have registered to
Yahoo!  and thus have a Yahoo! profile. One could expect that users
regularly query a set of tickers corresponding to stocks of their
interest. This is because
for queries that match
the ticker of a stock,
the search engine shows the user
up-to-date market information about the stock
in a separate display that appears
above the normal search results.
In addition, if any
important news appears, the corresponding page would show among the
top links in the search result. Therefore, we first compute the
distribution of the number of tickers searched by each user in various
time windows and time resolution (see Fig.  \ref{fig:tickers-per-user}).
Interestingly, most users search only one ticker, not
only within a month, but also within the whole year. This result is
robust along the time interval under observation and across
tickers. As a further step, among the users who search at least once a
given ticker in a certain time window, we compute the distribution of
the number of different days in which they search again for the same
ticker. In this case, we restrict the analysis to some specific
tickers, namely to those with highest cross correlation between query
volumes and trading volumes (e.g., those for Apple Inc., Amazon.com,
Netflix Inc.). Surprisingly, as shown in Section ``Materials and Methods'', Figs. \ref{fig:tickers-per-user}-\ref{fig:NFLX-activity}, the majority
of users ($\sim90\%$) searched the ticker only once, not only during a
month, but also within a year. Again, this result is robust along the
12 months in our dataset. Altogether, we find that most users search
for one ``favorite'' stock, only once. The fact that these users do
not check regularly a wide portfolio of stocks suggests that they are
not financial experts. In addition, there is no consistent pattern
over time. Users perform their searches in a seemingly uniform way
over the months.
In addition we find that our results are typical and very stable in time. In fact in this respect  
we do not observe any correlation between large fluctuations of trade volume, large price drops and influx of one-time searchers or with large price drops. In Fig. \ref{evoonetime} we show the evolution of one-time searchers which appears to be very stable in time. 

Overall, combining the evidence on the relation between query and trading
volumes with the evidence on individual user behavior, brings about a
quite surprising picture: movements in trading volume can be
anticipated by volumes of queries submitted by non-expert users, a
sort of \textit{wisdom of crowds} effect. 


\section*{Discussion}
In conclusion, we crawled the information stored in query-logs of the
Yahoo! search engine to assess whether signals in querying activity of
web users interested in particular stocks can anticipate movements in
trading activity of the same stocks.
Differently from previous studies we considered daily time series and
we focused on trading volumes rather than prices.\\
Daily volumes of queries related to a stock were compared with the
effective trading volume of the same stock by computing 
time-delayed cross-correlation. 

Our results show the existence of
a positive correlation between
todays stock-related web search traffic
and the trading volume of
the same stocks in the following days.
The direction of the correlation is
confirmed by several statistical tests. 

Furthermore, the analysis of individual users' behavior shows that
most of the users query only one stock and only once in a month. This
seems to suggest that movements in the market are anticipated by a
sort of "wisdom of crowd" \cite{easley2010connected}. These findings do
not explain the origin of the market movements but shows that that
search traffic can be a good proxy for them.

Furthermore, if one could assume that queries of a user reflect the
composition of her investment portfolio, our finding would suggest
that most of the investors place their investments in only one or two
financial instruments. The assumption that queries reflect portfolio
composition is a strong hypothesis and cannot be verified in our data
at the current stage. The finding would then deviate from the
diversification strategy of the well-known Markovitz approach, but
would be in line with previous empirical works on carried out on
specific financial markets.
This result, if confirmed, could have very important consequences. In
epidemics, by taking for granted that everybody has a mean
number of contacts brings to incorrect results on disease propagations. Here the assumption that 
investors portfolio is balanced, while it is not, could explain why  domino effects in the market are 
faster and more frequent than expected. \\
This does not mean that we can straightforwardly apply the models of 
epidemic spreading \cite{balcan2009mobility,romu1,romu2}  to financial markets. 
In fact, in the latter case (differently from ordinary diseases) panic spreads mostly by 
news. In an ideal market, all the financial agents can become ``affected'' at the same time by the 
same piece of information.   
 This fundamental difference makes the typical time scale of reactions in financial markets much 
shorter than the one in disease spreading. 
 It is exactly for that reason that any early sign of market behavior 
must be considered carefully in order to promptly take  the necessary countermeasures. 
 We think that this information can be effectively used in order to detect early signs of financial 
distress. 

We also believe this field to be very promising and we are currently working on the extension of this 
kind of web analysis to twitter data and semantic analysis of blogs.


\section*{Materials and Methods}
In this section we give a detailed overview of the investigations
carried out in this paper. The first contribution of our work
consists, as previously said, of an analysis of the relation between
the activity of the users of the Yahoo! search engine and real events taking place within the stock market. 
Our basic assumption is that any market activity in an individual
stock may find some correspondence in the search activity of the users interested in that stock. 
Thus we study whether significant variations in the stock trading volumes are anticipated by analogous variations in the volume of related Web searches.
To investigate the existence of a correlation between query volumes and trading volumes, we compute time-lagged \textit{cross-correlation coefficients} of these two series. 

We conduct such analysis performing separate experiments to test the
two different query definitions that we take into consideration, i.e.,
queries containing the stock ticker string, or queries matching the company name.  The results of this first set of experiments are presented in Subsection ``Correlation between query volumes and trading volumes''. 

We then apply permutation tests, Granger-causality test and several analyses to assess the significance of the correlations found. These experiments are described in Subsection ``Statistical validation of query anticipation''.

Finally, Subsection ``Analysis of users' behavior'' presents details
of the last part of our work, where we try to gain a better knowledge of the typical behavior of the users who issue queries related to finance. 
Here we refine our analysis of the information extracted from query logs to understand what a typical user searches for, such as whether she looks for many different tickers or just for a few ones, and, if she looks for them regularly or just sporadically. 

\subsection*{Database} 
\subsubsection*{The stocks analyzed}\label{sec:data-stocks}
In this work we compare query volumes and trading volumes of a set of companies traded in the NASDAQ 
(National Association of Securities Dealers Automated Quotation) 
stock exchange, which is the largest electronic screen-based equity securities trading market in the United States and second-largest by market capitalization in the world.
Precisely, we analyze the  $100$ companies  included in the NASDAQ-100 stock-market capitalization index.
These companies are amongst the largest non-financial companies that are listed on the NASDAQ 
(technically the NASDAQ-100 is a modified capitalization-weighted index, it does not contain financial companies and it also includes companies incorporated outside the United States.) We list these companies  in Table \ref{table:1}. 
The daily financial data for all of stocks is publicly available from Yahoo!\ Finance\footnote{http://finance.yahoo.com/} and we focus our attention on the daily trading volumes.
\subsubsection*{Query data}\label{sec:data-queries}
The query-log data we analyze is a segment of the Yahoo! US
search-engine log, spanning a time interval of one year,
from mid-2010, to mid-2011. 
The query-log stores information about actions performed by
users during their interactions with the search engine, including the queries they submitted and the result pages they were returned, as well as the specific documents they decided to click on.

We compute query volume time series by extracting and aggregating on a daily
basis two different types of queries for each traded company:
\begin{itemize}
\item all queries whose text contains the stock ticker string (i.e. ``YHOO'' for Yahoo!) as a distinct word;
\item all queries whose text exactly matches the company name (after
  removing the legal ending, ``Incorporated'' or ``Corporation'' or ``Limited'', and all their possible abbreviations).
\end{itemize}

All queries in the log are
associated with a timestamp
that represents the exact moment the query was issued to the search engine.
We use this temporal information to aggregate the query volumes
at different levels of granularity. 
Furthermore, every action is also annotated with a cookie, representing the 
user who submitted the query. These cookies allow to track the activity of a single user 
during a time window of a month.
By using this information, we also computed \emph{user volumes} by counting the daily number of distinct users who made at least one search related to one company 
(according to the query definitions provided above). 
Thus, for each stock taken into consideration, we can compare the
daily volumes of related queries, as well as
the number of distinct users issuing such queries per day
with the daily trading volumes gathered from Yahoo!\ Finance.



\subsection*{Correlation between query volumes and trading  volumes}
\label{sec:correlations}
We compare the query volume
of every stock with the trading volume of the same stock. 
The two definitions of queries introduced are used in separate experiments, 
that is, in one case we aggregate all the queries containing the
ticker of a company, and in another case we only consider queries that
match the company name. 

We extract from both data sources (the query volumes and the trading
volumes of a given stock) a time series composed by daily values in
the time interval ranging from mid 2010 to mid 2011. 
Although the query-log contains information collected during holidays
and weekends as shown in Fig. \ref{fig:AAPL-days} for the case of the
AAPL stock, the financial information is obviously only available for
trading days. Thus, for the sake of uniformity, we filter out all the
non-working days from the query volume time series.
In the end, we obtain two time series of 250 working days for every
stock.

As a second step, given the time series $Q$ of the query volumes and the time series $T$ of trading volumes, we compute the \textit{cross-correlation coefficient}
$r(\delta)$ for every company.\\
This correlation coefficient ranges from $-1$ to $1$. 
Although the above coefficient can be computed for all delays $\delta=0,1,\dots,N$, we chose to consider a maximum lag of one week (five working days). \\

Tables \ref{tbl:nasdaq-ticker} and~\ref{tbl:nasdaq-companies} report the results obtained for these experiments. Columns instead correspond to different values of the time-lag $\delta$ used in the calculation of the cross-correlation coefficients. We observe that the cross-correlation coefficients always assume nearly equal to zero for $|\delta|>5$. 

When the first query definition is taken into consideration (ticker query), the average cross-correlation coefficient in the base case of $\delta=0$ is equal to $0.31$.
Similar values are obtained if a time-lag $\delta$ in the range $[-2,2]$ is considered. It is worth noticing that for some individual companies we observe much higher correlations. On this account Table \ref{tbl:best} presents the best results for single stocks (see Tables S1 and S2 of Supporting Information for the complete results: it is worth noticing that considering only the stocks for which $r(1)>0$, there are 8 stocks for which $r(1)< 0$, for 68 stocks it holds that $r(1)>r(-1)$ while for the remaining 11 stocks we observe $r(1)\leq r(-1)$).
For these companies, we also report in Table \ref{tbl:drop} (Tables S4 and S5 for all the results) the basic cross-correlation at lag $\delta=0$ after removing from the time series the days corresponding to the top $5\%$
and $10\%$ values of the trading volume. It is interesting to observe that the correlations are still significant and so the correlation does not seem to be due only to peak events, 
which generally correspond to headlines in the news, product announcements or dividend payments.

When the second query definition (company names) is considered, we observe weaker correlations than the previous case. The average cross-correlation coefficient in the base case $\delta=0$ is equal to $0.12$.

In addition we point out that the process of extracting data from
query-logs can introduce spurious queries which have a non financial origin.
Especially some of the ticker queries match our above definition,
but are nonetheless unrelated to the stock represented by the ticker.
For instance, some ticker strings correspond to natural language
words, such as ``FAST'' (Fastenal Company) and ``LIFE'' (Life Technologies Corp.).
As one can reasonably expect, the overwhelming
majority of queries containing these words are completely unrelated to
the companies that are the subject of our study. Other cases of
companies for which we discovered very large levels of noise included
e-commerce portals like Ebay. In all these cases the ticker often
appears in navigational queries that are unrelated to the company stock (see Table S3 of Supporting Information).
For this reason, we filter out all companies whose query volumes are discovered to be 
noisy, retaining a smaller, but cleaner set of $87$ companies for which the spurious queries are a negligible fraction.
By restricting the computation of the cross-correlation function to these companies, we observe a
larger value of the average cross-correlation. Table \ref{tbl:clean-ticker} reports the results obtained for the first query definition (queries including the ticker as a distinct word), which represents the case for which the best performances of the queries are observed. The average cross-correlation at time lag $\delta=0$ is $0.36$.

Besides query volumes, we also consider user volumes, i.e., the number
of distinct users who issued queries related to a company in any given
day. For reasons listed above, this analysis is restricted to the 87
NASDAQ-100 companies for which we have a clean query-log signal. 
Cross-correlations between user volumes and trading volumes
are shown in Table \ref{tbl:clean-user}. 
We observe similar findings to the ones obtained in the previous
experiments, although the average cross-correlation is $5\%$ smaller
than the one obtained with query volumes.
The average cross-correlation
between user volumes and trading volumes
at time lag $\delta=0$ is $0.31$.
\subsection*{Statistical validation of the query anticipation}
\subsubsection*{Permutation test}\label{sec:permutation}
A permutation test, also called randomization test, is a statistical significance test where 
random rearrangements (or permutations)  of the data are used to
validate a model.
Under the null hypothesis of such a test
data permutations have no effect on the outcome,
and the reshuffled data present the same properties 
as the true instance.
The rank of the real test statistic among the shuffled test statistics determines 
the empirical ``p-value'', which is the probability that the test statistic would be at least as extreme as observed, if the null hypothesis were true. 
For example, if the value of the original statistic is $95\%$ greater than the random values, we can reject the null hypothesis with a confidence $p < 0.05$.
This means that the probability that we would observe a value as
extreme as the true one, if the null hypothesis were true, is less
than $5\%$.
In our setting, the aim is to verify the significance of the
correlation between the queries containing the ticker of a company and
the trade volumes of the same company. 
In particular,
we want to assess if the cross-correlation
between query volume and trading volume of a given company
is higher than the cross-correlation
between query volume of company $C_i$
and trading volume of some other company $C_j \neq C_i$.
The purpose of this test is to show that
the correlations we observe are
not merely a consequence of
stock market related web search activity being
correlated with stock market activity {\em in general}.


Our original data is given by the set of pairs of time series $\{ Q_i,T_i\}$ 
previously considered. 
Every pair in this set contains information concerning a given company $C_i$. As already indicated, 
$Q_i$ is the time series of the query volumes of $C_i$, whereas $T_i$ is the time series of the trading volumes of $C_i$. 
We use as test statistic the cross-correlation coefficient between $Q_i$ and $T_i$. 
Starting from the above data, we apply 1000 random permutations to
create an ensemble of 1000 distinct datasets, each one composed of
pairs $\{ Q_i,T_j\}$, where the time series of query volumes of a company $C_i$ is randomly paired with the time series of trade volumes of a different company $C_j$. For each pair $\{ Q_i,T_j \}$ included in each randomly generated dataset,
we compute the cross-correlation between $Q_i$ and $T_j$. 

We then compare the (macro-)average cross-correlation that we get for the real data with the average values obtained for the 1000 randomized datasets in which the queries of a company
are always paired with the trades of another company. While the average result that we get for the original data is  $\langle r^{Original}(0)\rangle=0.31\pm 0.05$, the values obtained for the 
test statistic when the random permutations are applied are much smaller. We find $\langle r^{Reshuffled}(0)\rangle \in[-0.033,0.06]$. 
Therefore we get an empirical p-value of 0.001, meaning that the correlations observed on the real data are statistically significant at $0.1\%$.

We also check the significance of the correlations obtained for individual companies separately. Our goal here is to understand on a deeper level what companies are
actually correlated with the corresponding queries, and which ones are not. We consider the two scenarios below.

\begin{enumerate}
\item  In the first case, the null hypothesis is the following:  
{\em The correlation between trading volume of company $C_i$ and
 query volume of the same company
 is not higher than the correlation between trading volume of company
 $C_i$ and query volume of some other company $C_j$.}
For every company $C_i$, we compare the real data $\{Q_i,T_i\}$ with
the 1000 $\{Q_j,T_i\}$ pairs where each $Q_j$ comes from one of the 1000 random datasets generated before.
The test statistic that we use for the comparison is the same as before, that is, the cross-correlation coefficient $r(\delta)$
between the two time series forming any given pair. 
For every company $C_i$, we compute the empirical p-value by taking the rank of the real test statistic $\langle r^{Original}(0)\rangle$ within the sorted order of the values computed from reshuffled data.
 
\item  Similarly, in the second scenario, our null hypothesis is:
{\em The correlation between query volume of company $C_i$ and
 trading volume of the same company
 is not higher than the correlation between query volume of company
 $C_i$ and trading volume of some other company $C_j$.}
Now, for any query-volume $Q_i$, the real data is still given by the
pair $\{Q_i,T_i\}$. We compare this with the 1000 $\{Q_i,T_j\}$ pairs
where each $T_j$ comes from a different random dataset. We calculate
the cross-correlation between the two time-series included in every
pair, and determine the p-values in the same way as above. 

\end{enumerate}

In both the scenarios taken into consideration, for most of the companies the test rejected $H_0$. More specifically,

\begin{enumerate}
\item: We got the minimum p-value $(0.001)$ for 50 companies (out of 87). The p-value was $ \ge  0.05$ in 19 cases.
\item: We got the minimum p-value $(0.001)$ in  48 cases. The p-value was  $ \ge 0.05$ in 26 cases.
\end{enumerate}

To summarize, we observe that for $3/4$ of the stocks the correlation
between query volume and trading volume
can not be explained by a simple global correlation
between finance related search traffic and market activity in general.

It is worth noting that large p-values are related to companies for which poor correlation is present between query-log data and trading, maybe because of the large noise in the dataset.

\subsubsection*{Correlation between query volume and volatility}
Trading volume and volatility are correlated and volatility is autocorrelated. Therefore a source of the correlation between present query volume and future trading volume can be the autocorrelation component of volatility. Here we show that the origin of these correlations cannot be traced back to volatility. In order to perform such a task we compare the correlation 
between query volume and absolute price returns (i.e a proxy for the volatility) with the one between query volume and trading volume.\\
We define the \textit{price return} of a day $t$ as follows: 
$$  R(t) =  P_c(t) - P_c(t-1)  $$
where $P_c$ is the closing price of the day $t$.
For each stock in our NASDAQ-100 \textit{clean} list we compute the price returns and build three time series: 

\begin{itemize}
\item The time series $\mathbf{P_A}$ of the unsigned price returns:
  $P_A = \{  | R(t) | : t = 2, \dots, N \}$
\item The time series $\mathbf{P_+}$  of the positive price returns: $P_+ = \{  R(t) :  t = 2, \dots, N  \mbox{  } s.t. \mbox{ } R(t) > 0 \}$
\item The time series $\mathbf{P_-}$  of the negative price returns:  $P_- = \{ R(t) : t = 2, \dots, N  \mbox{  } s.t. \mbox{ } R(t) < 0\}$
\end{itemize}
The time series $P_A$ of the unsigned price returns has $N - 1$ elements, being $N$ the length (number of days) of the time interval covered by our data ($N=250$). 

Similarly to the experiments involving trading volumes, we compute for
every stock the cross-correlation $r(\delta)$ between the price returns and the query volume of the same company.

\begin{figure}[!ht]
\begin{center}
\includegraphics[scale=0.4]{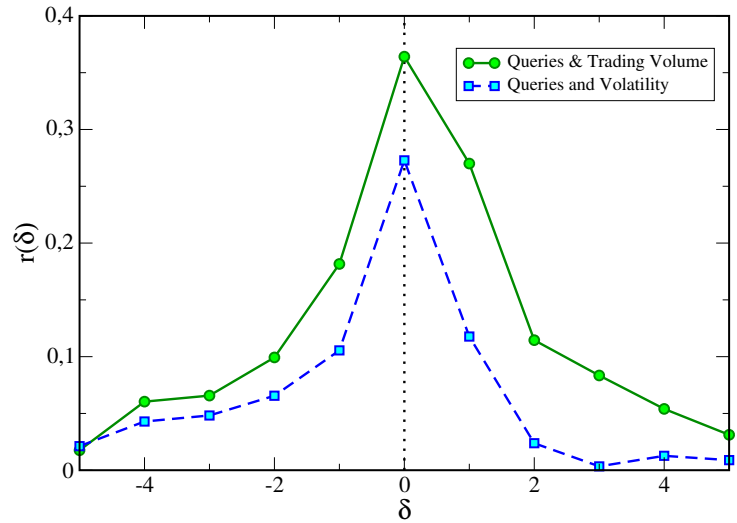}
\end{center}
\caption{\textbf{Comparison of the cross-correlation function between query volumes and trading volumes and query volumes and volatility}. Trading volume and volatility are correlated and given the fact that volatility is also autocorrelated, the correlation between present query volume and future trading volume could be simply originated by this autocorrelated term. However, we show that the cross-correlation between query and volatility (broken line) is significantly smaller than the one between query and trading volume (solid line). Moreover the $\delta>0$ branch in the volatility case is equal or even smaller than the value observed in the $\delta<0$ one. If the origin of the effect were due to the autocorrelation component of the volatility, we would expect a similar behavior for both cross-correlation function. This facts support that the non-autocorrelated origin of the correlation between between present query volume and future trading volume.
As a proxy for the volatility we use the absolute value of daily price returns.}
\label{figure3}
\end{figure}

\begin{figure}[!ht]
\begin{center}
\includegraphics[scale=0.4]{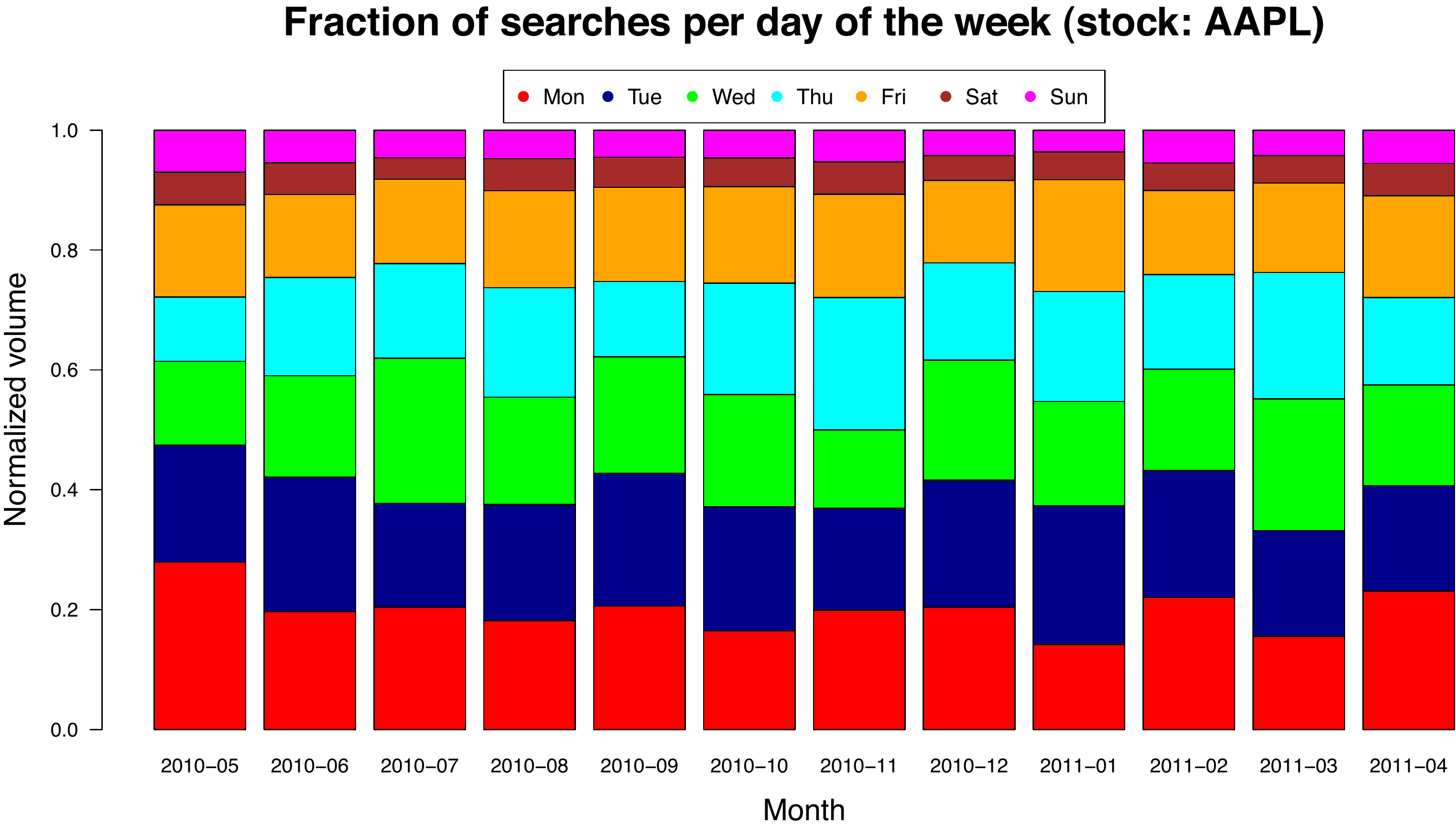}
\end{center}
\caption{\textbf{Query-search  for \textbf{AAPL} stock in the various days of the week}. Query volumes of NASDAQ-100 tickers  are negligible during non-working days, then  we consider only the contribution to query volumes deriving from working days.  
}
\label{fig:AAPL-days}
\end{figure}

\begin{figure}[!ht]
\begin{center}
\includegraphics[scale=0.4]{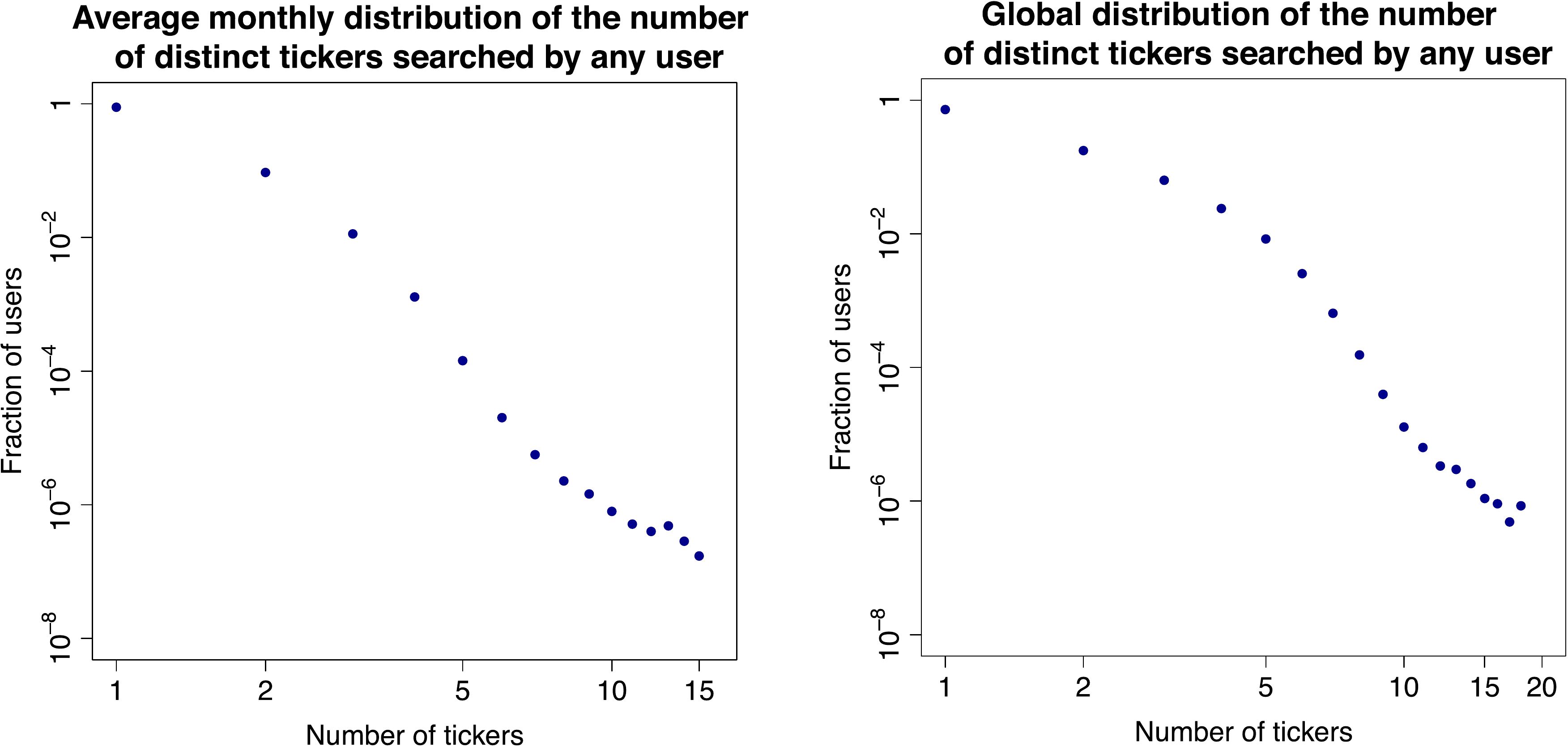}
\end{center}
\caption{\textbf{Typical users' behavior.} Average (left)  monthly and (right) yearly distribution of the number of distinct tickers searched by any Yahoo! user.  
}
\label{fig:tickers-per-user}
\end{figure}

\begin{figure}[!ht]
\begin{center}
\includegraphics[scale=0.4]{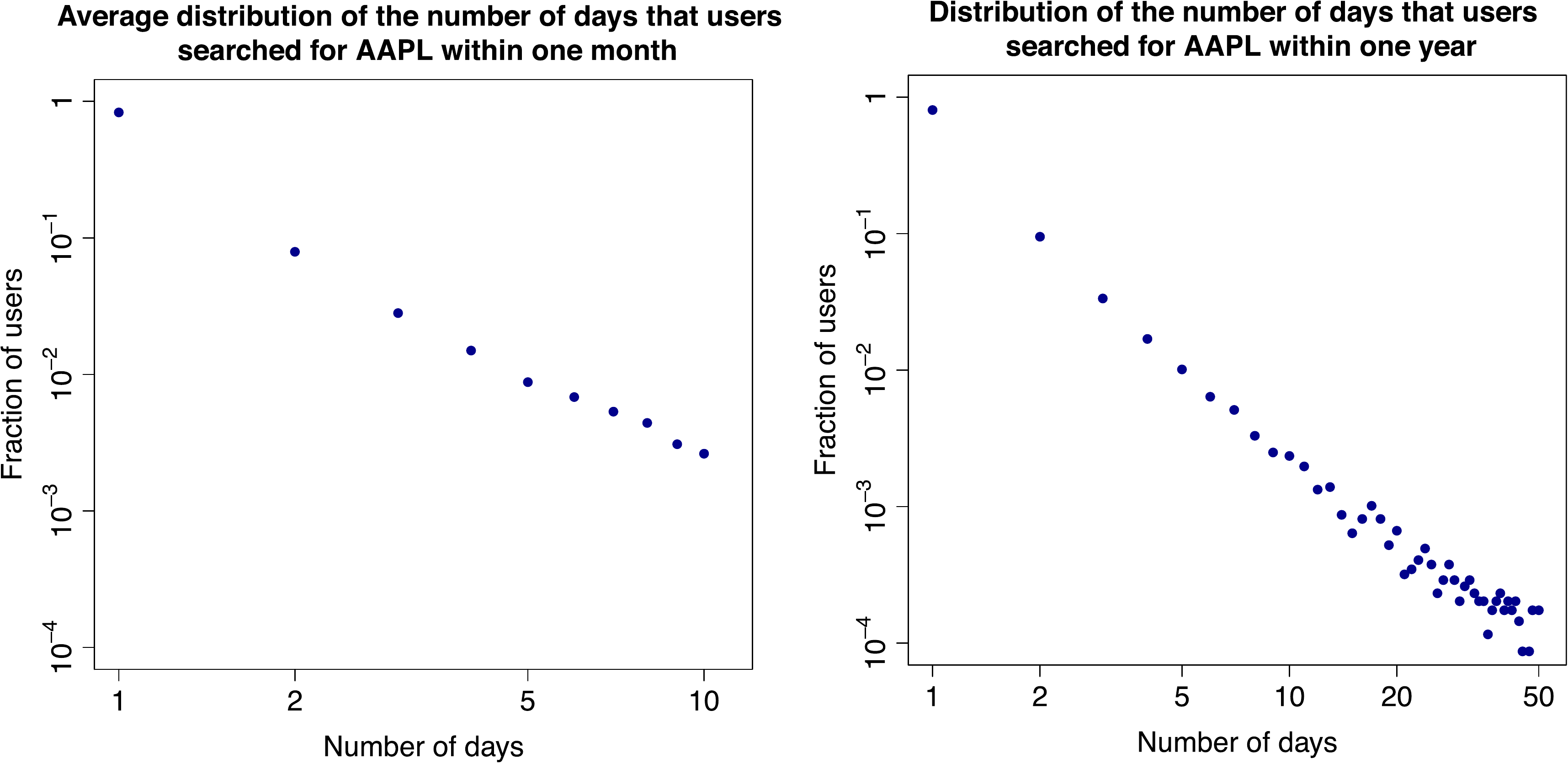}
\end{center}
\caption{\textbf{Behavior of the users who search for AAPL.} Distribution of the number of days that users searched for  AAPL within one month (left) and over the whole year (right).  
}
\label{fig:AAPL-activity}
\end{figure}

\begin{figure}[!ht]
\begin{center}
\includegraphics[scale=0.4]{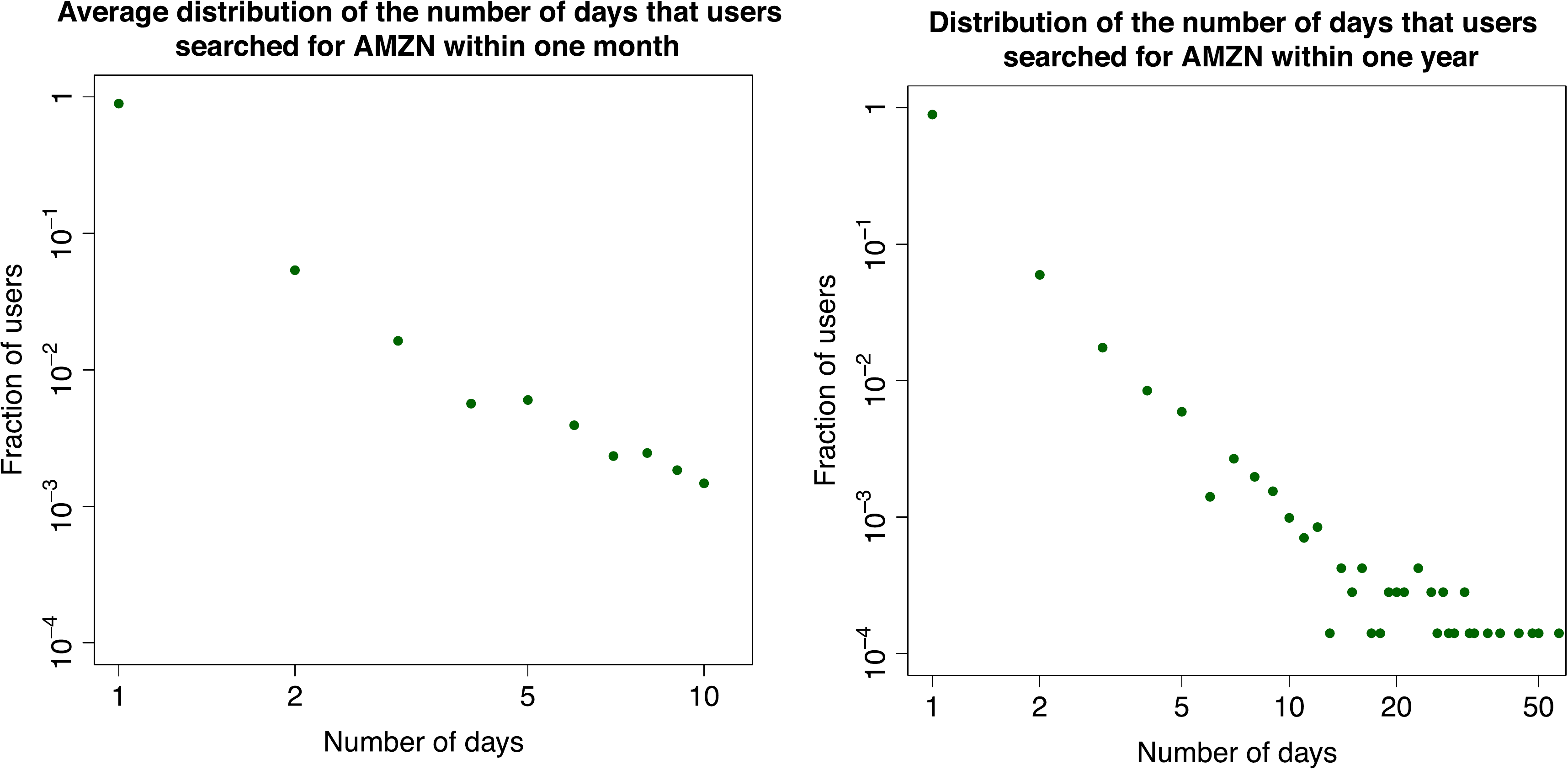}
\end{center}
\caption{\textbf{Behavior of the users who search for AMZN.} Distribution of the number of days that users searched for  AMZN within one month (left) and over the whole year (right).  
}
\label{fig:AMZN-activity}
\end{figure}

\begin{figure}[!ht]
\begin{center}
\includegraphics[scale=0.4]{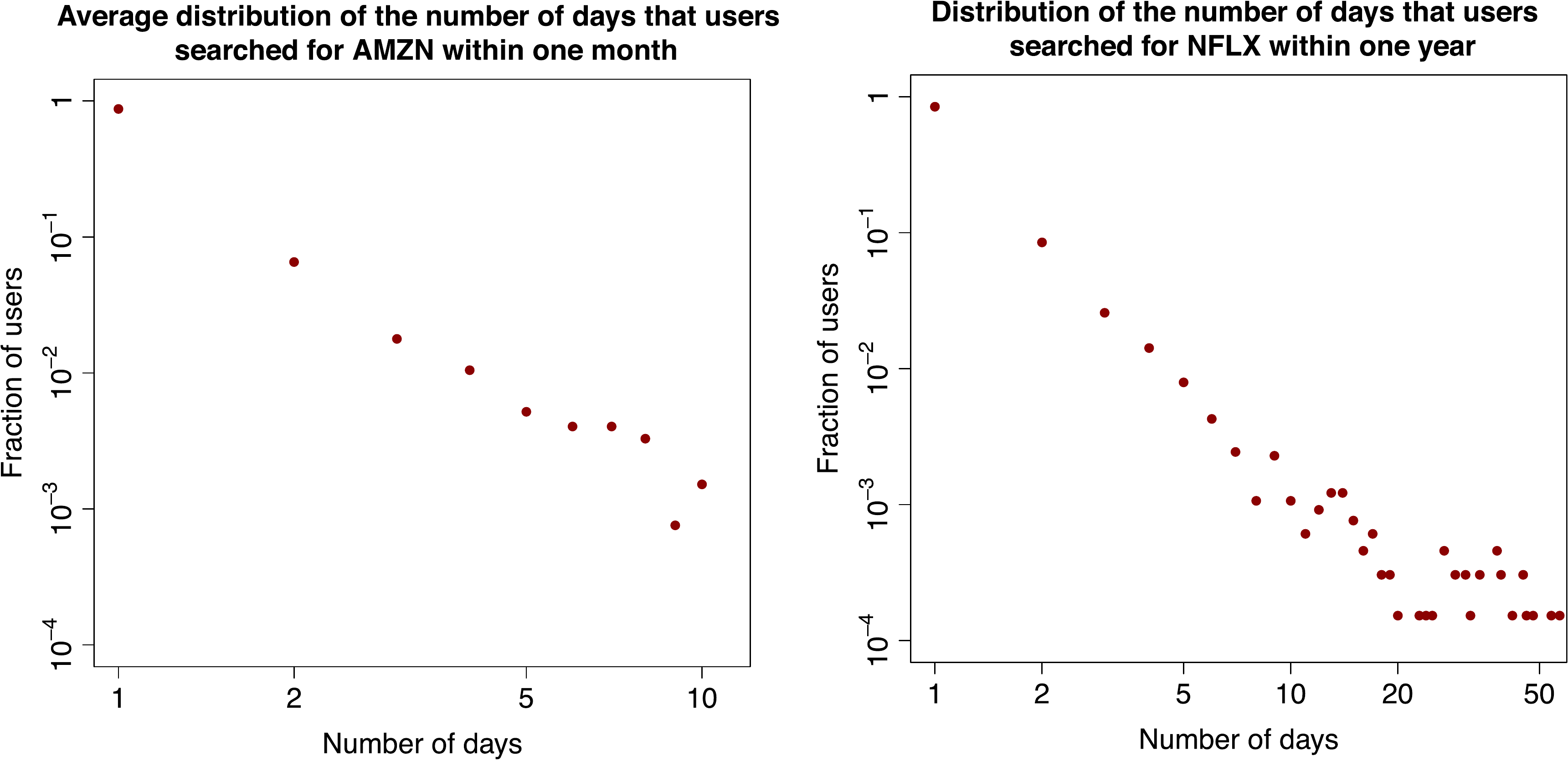}
\end{center}
\caption{\textbf{Behavior of the users who search for NFLX.} Distribution of the number of days that users searched for  NFLX within one month (left) and over the whole year (right).  
}
\label{fig:NFLX-activity}
\end{figure}

\begin{figure}[!ht]
\begin{center}
\includegraphics[scale=0.4]{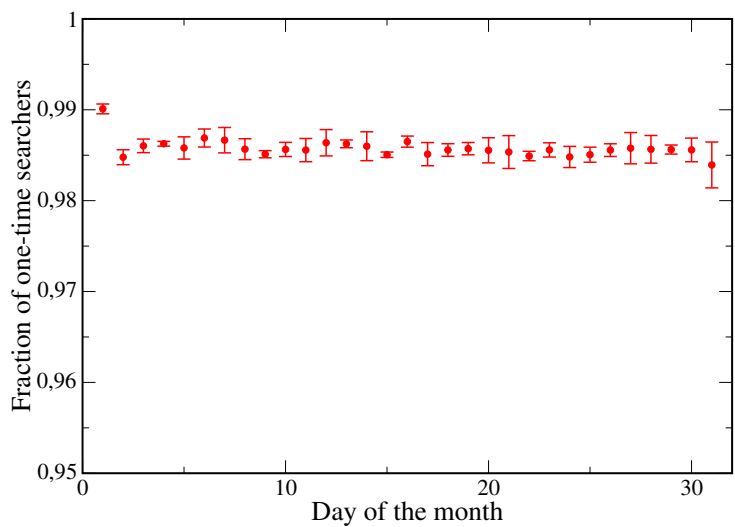}
\end{center}
\caption{\textbf{Evolution of the percentage of one-time searchers}. The fraction of one-time searchers appear to be very stable in time and we do not observe a correlation of these kind of users with anomalous trading volume or price movements. }
\label{evoonetime}
\end{figure}

Fig. \ref{figure3} (broken line) reports the cross-correlation
function between the unsigned price returns and query volume.
The average value of the basic cross-correlation at lag $\delta = 0$ between query volume and price returns is $0.2728$. 
This result reflects the fact that
in days when
the prices of the NASDAQ-100 stocks exhibit a large variation (either
positive or negative), there is a considerable amount of web search activity concerning the same stocks.

However, as shown in Fig. \ref{figure3} the cross-correlation between
query volume and volatility (broken line) is significantly smaller
than the one between query volume and trading volume (solid
line). Moreover the $\delta>0$ branch in case of volatility is equal or even smaller than the value observed in the $\delta<0$ one. If the origin of the effect were due to the autocorrelation component of volatility, we would expect a similar behavior for both cross-correlation functions. These facts support the non-autocorrelated origin of the correlation between between todays query volume and future trading volume.
 

For the time series $P_+$ (positive returns) and $P_-$ (negative
returns), we only computed the cross-correlation between query volumes for lag
$\delta = 0$. The reason is due to the fact that
the time gap between two consecutive elements of those series is variable.
The average correlations obtained for the $87$ clean NASDAQ tickers are report in Table~\ref{tbl:signed-price-returns}. 
The results are similar to ones we get for the unsigned price returns.

\subsubsection*{Granger Causality}

\label{sec:granger}

The Granger-Causality test is widely used in time-series analysis to determine whether a time series 
$X(t)$  is useful in forecasting another time series $Y(t)$. 
The idea is that if  $X(t)$ Granger-causes $Y(t)$ if $Y(t)$ can be better predicted using both the histories of $X(t)$ and $Y(t)$ rather than using only the history of $Y(t)$.  
The test can be assessed by regressing $Y(t)$ on its own time-lagged values and  on those of $X(t)$. 
 An F-test  is then used to examine if the null hypothesis that $Y(t)$ is not Granger-caused by 
 $X(t)$ can be rejected. 

In this work, we apply the Granger-causality test to analyze the relation between query volumes and trading volumes, and also between user volumes and trading volumes.
Our aim is to prove that search activity related to a company,
Granger-cause the trading volume on the company stock. However, we also want to verify whether the notion of Granger causality holds in the opposite direction. Hence, we apply the test in the two possible directions.

Again, we first consider all companies included in the NASDAQ-100 data set. However, given that we know from the previous analysis that in some cases the query volumes are very noisy and not related to the traded company they have been extracted for, we also perform the test on the smaller test of $87$ companies obtained through manual filtering. 

Table \ref{tbl:granger} presents the results of the Granger-causality test. 
Each row in the table summarizes the outcome of an experiment. The table specifies the two available query-log time series (query volumes Q or user volumes U) compared with trading volume T (comparisons are always made for each company independently), the lag applied (expressed in terms of number of days), the direction in which the test is applied : $X \rightarrow Y$ means that the null hypothesis $H_0$ is ``$X$ does not Granger-cause $Y$''. The last three columns provide a summary of the results obtained for all companies that are taken into consideration during the test. The fourth and fifth column respectively report the percentage of companies for which the null hypothesis was rejected with $p < 0.01 (0.05)$. The last column reports the average reduction in RSS.

In all the cases, it can be observed that the $ \rightarrow T$
direction of the test
is much stronger than the opposite direction $T
\rightarrow . $.
That is, we obtained stronger support for the case that
time-series extracted from the query-log Granger-cause the trading
volume of the same company, as opposed to
trading volume Granger-causing query or user volumes.
Especially this is the case when significance at 1\% is required.

For instance, let us consider rows 9 and 11 in the Table \ref{tbl:granger}. When the \emph{clean} set of $87$ tickers is examined, we observe that in $45.35\%$ of the cases the null hypothesis ($Q$ does not Granger-cause $T$) is rejected with $p < 0.05$, and for $33.72\%$ of the companies the same held with with $p < 0.01$. A much weaker result is obtained when the opposite direction is considered. Only for $5.8\%$ of the companies the null hypothesis could be rejected with  $p < 0.01$. 

As we have already observed in the cross-correlation experiment, we
get slightly weaker results when considering user volumes. Observe
line 11 in the table: in $29.1\%$ of the cases the trading volume $T$
is Granger-caused by the user volume $U$ with probability greater than $99\%$. The average reduction in RSS is $4\%$. 

 In short, adding information about todays query volume reduces the average prediction error (in an autoregressive model) for tomorrows trading volume by about $5\%$.
For half of the companies the reduction is statistically significant
at $1\%$, that is, both query volume and user volume Granger-causes the trading volume. We can also interpret this as follows: 
query/user volume helps to predict the trading volume, but the reverse does not hold. 

It can be now argued that the Granger test, in principle, should be
used only on series for which the error term in the regressions is
gaussian. In this framework instead we are dealing with fat-tailed
distribution underlying the query volume and trade volume series (see
Figs. S1-S6 of Supporting Information). However, in the next section we present a
series of analyses which confirm the significance of the results found
here. In particular, they all support the evidence that todays web
search traffic is more informative on tomorrows trading activity than the reverse case.  

\subsubsection*{Beyond Granger Causality}
To study the anticipation effect and the power of search engine data for predicting stock trading volumes, we performed several statistical tests checking various hypotheses. 
The tests are detailed below. 
\paragraph{Test 1} 
To test if query volume can predict future trading volume, denoted $Q \rightarrow T$, we use four different regression models:

\begin{enumerate}
\item $M_1: T_t \sim T_{t-1}$: \\ We predict trading volume of tomorrow using trading volume of today.
\item $M_2: T_t \sim T_{t-1}, Q_{t-1}$: \\ We predict trading volume of tomorrow using both trading and query volume of today.
\item $M_3: Q_t \sim Q_{t-1}$: \\ We predict query volume of tomorrow using query volume of today.
\item $M_4: Q_t \sim T_{t-1}, Q_{t-1}$: \\ We predict query volume of tomorrow using both trading and query volume of today.
\end{enumerate}

Let $R^2(M_i)$ denote the sum of squared residuals for model $M_i$. 
We define $$ \Delta( Q \rightarrow T ) = R^2(M_2) - R^2(M_1) \mbox{,  and } \Delta( T \rightarrow Q ) = R^2(M_4) - R^2(M_3). $$
In other words $\Delta( Q \rightarrow T )$ is the variation of $R^2$ when we use $Q_{t-1}$ to predict $T_t$ \textit{in addition} to $V_{t-1}$.
Likewise, $\Delta( T \rightarrow Q )$ is the variation in $R^2$ when $T$ is added to an auto-regressive model of $Q$. 

Our aim is to test the following hypotheses: 
\begin{enumerate}
\item Null-hypothesis $\mathbf{H_0}$:  
$\Delta( Q \rightarrow T )$ and $\Delta( T \rightarrow Q )$ \textit{are not significantly different}.

\item Alternative hypothesis $\mathbf{H_1}$ :
$\Delta( Q \rightarrow T )$  \textit{is significantly larger than} $\Delta( T \rightarrow Q )$.

\item Alternative hypothesis $\mathbf{H_2}$:  $\Delta( T \rightarrow Q )$ \textit{is significantly larger than} $\Delta( Q \rightarrow T )$.
\end{enumerate}
To compare $\Delta( Q \rightarrow T )$ and $\Delta( T \rightarrow Q )$, we apply a bootstrap procedure to estimate their distribution. 
We generate $9\,999$ samples for $\Delta( Q \rightarrow T )$ and $9\,999$ samples for $\Delta( T \rightarrow Q )$, using the \textit{case resampling} strategy. 
We denote by $\Delta_{bs}( Q \rightarrow T )$ the bootstrap distribution of $(R^2(M_2)-R^2(M_1))$, and by $\Delta_{bs}( T  \rightarrow  Q )$ the bootstrap distribution of $(R^2(M_4)-R^2(M_3))$.\\
Given $\Delta( Q \rightarrow T )$ and $\Delta_{bs}( T \rightarrow Q )$, we can derive an empirical p-value of $\Delta( Q \rightarrow T )$ being larger than $\Delta( T \rightarrow Q )$. 
This p-value, which we denote by $p(Q \rightarrow T)$, is computed as the the rank of $\Delta( Q \rightarrow T )$ in the list of sorted $\Delta_{bs}( T \rightarrow Q )$ values divided by $n+1$, where $n$ is the number of bootstrap samples. 
Depending on the chosen significance level, by the empirical p-value we can now reject $H_0$, and support $H_1$.


We run this test for the list of  \textit{clean} NASDAQ-100 tickers.
For 26  companies we obtain an empirical p-value lower than $0.01$: this result suggests that, for these companies, we can reject the null hypothesis at the significance level of $0.01$, finding support for $H_1$.\\
Table~S6 (see Supporting Information) reports the list of these companies, together with the respective p-values $p(Q \rightarrow T)$ and $p(T \rightarrow Q)$. The third column of the table contains the value 
of the basic cross-correlation at lag $\delta = 0$ between query volume and trading volume. 

We also test the opposite direction. To verify if there is  any support for $H_2$, we took 
 $\Delta( T \rightarrow Q )$ and $\Delta_{bs}( Q \rightarrow T )$, and use the same procedure as above to compute the  empirical p-value of $\Delta( T \rightarrow Q )$ being larger than $\Delta( Q \rightarrow T )$. 
This time, all p-values $p(T \rightarrow Q)$ that we obtain for the 87 clean tickers are very large.
In almost every case $\Delta( T \rightarrow Q )$ is smaller than the values in $\Delta_{bs}( Q \rightarrow T )$.
This suggests that trading volumes of today do not help in \textit{predicting} query volumes of tomorrow.

In Table~S7 we report the ten tickers with the smallest $p( T
\rightarrow Q )$: observe that even the smallest values are much
larger than $0.01$, thus we not find any convincing support for $H_2$. 

\paragraph{Test 2} 
The previous test is based on the idea of comparing the improvement in
$R^2$ after adding information from the second time series to an auto-regressive model. The test that we present below is based on the direct comparison of the $R^2$ values of $Q  \rightarrow  T$ and $T  \rightarrow  Q$. 

We consider the two following regressive models: 
\begin{enumerate}
\item
$ M_1:  Q_t \sim T_{t-1}$
\item 
$ M_2:  T_t \sim Q_{t-1}$
\end{enumerate}

We perform the two regressions above, and compute the respective $R^2$ values, which we call $R^2(T \rightarrow Q)$ and $R^2(Q \rightarrow T)$.
If   $R^2(T \rightarrow Q) \ge R^2(Q \rightarrow T)$, then we conclude $Q  \rightarrow  T$, and viceversa.

To assess the significance of the test, we generate $1\,000$ bootstrap vectors starting from the real data and applying random sampling with replacements. 
We compute $M_1$ and $M_2$ on the bootstrap vectors, obtain the corresponding residuals, and extract the $95$-th percentiles $R^2_{95}(T \rightarrow Q)$ and $R^2_{95}(Q \rightarrow T)$, that is, the values such that, for $95\%$ of the boostrap vectors, the sum of squared residual is below this values. 
Then we compare  $R^2(T \rightarrow Q)$  with  $R^2_{95}(Q \rightarrow T)$, and $R^2(Q \rightarrow T)$  with  $R^2_{95}(T \rightarrow Q)$. \\
We run this test on the clean set of NASDAQ-100 tickers. For a significance level of $0.05$, the outcome is the following: 
\begin{itemize}
\item 61 companies with a \textit{significant} difference at $p=0.05$ between  $\Delta(Q \rightarrow T)$  and $\Delta(T \rightarrow Q)$ values: 
 $55$ support $Q \rightarrow T$, and $6$ support $T \rightarrow Q$ (These are: joyg, lltc, rost, teva, vrsn, vrtx).
\item 26  companies have no \textit{significant} difference between the two directions (see Table S8 and S9 of Supporting Information).
\end{itemize}

\paragraph{Test 3}  

In this test we again consider the four regression models that are used for the first test: 
\begin{enumerate}
\item $M_1: T_t \sim T_{t-1}$: \\ We predict trading volume of
  tomorrow using the trading volume of today.
\item $M_2: T_t \sim T_{t-1}, Q_{t-1}$: \\ We predict trading volume of tomorrow using both trading and query volume of today.
\item $M_3: Q_t \sim Q_{t-1}$: \\ We predict query volume of tomorrow
  using the query volume of today.
\item $M_4: Q_t \sim T_{t-1}, Q_{t-1}$: \\ We predict query volume of tomorrow using both trading and query volume of today.
\end{enumerate}

We consider the following hypothesis: 
\begin{enumerate}
\item Null-hypothesis $\mathbf{H_0}$:  
$\Delta( Q \rightarrow T ) = 0$ 

\item Alternative hypothesis $\mathbf{H_1}$ :
$\Delta( Q \rightarrow T ) > 0$.

\end{enumerate}
To test if $Q  \rightarrow  T$, we compute the regression models $M_1$ and $M_2$, and  derive the corresponding residuals  $R^2(M_1)$ and $R^2(M_2)$. 
We then compute  $9\,999$ bootstrap estimates of $R^2$ both for $R^2(M_1)$ and $R^2(M_2)$. 
Next we compare these two bootstrap samples by applying the Mann-Whitney U test, also known as the Wilcoxon rank-sum test. 

The test is aimed at assessing whether one of two samples of independent observations tends to have
larger values than the other. It is based on the null-hypothesis of the two samples having equal medians.\\
We also test the opposite direction $T  \rightarrow  Q$. 
We compute the regression models $M_3$ and $M_4$, and the corresponding residuals  $R^2(M_3)$ and $R^2(M_4)$. 
We compute  $9\,999$ bootstrap estimates of $R^2$ both for $R^2(M_3)$ and $R^2(M_4)$, and we apply again the  Mann-Whitney U test.
For the 87 clean NASDAQ-100 tickers, we get the following results (see Table S10 of Supporting Information): 
\begin{itemize}

\item Only 3 out of 87 clean Nasdaq tickers are not significant at $p=10^{-4}$ when testing for $Q \rightarrow V$. These are LINTA $(p = 0.031)$, CHKP $(p = 0.034)$ and FISV $(p = 0.054)$. 
\item In the other direction, $V \rightarrow Q$, only 19 tickers are not significant at $10^-4$.
\item In every other case the p-value is approximately $~0$. This might be due to the Mann-Whitney test being better suited for small sample sizes.

\end{itemize}

\subsection*{Analysis of users' behavior}

\label{sec:user-behavior}

We now investigate the typical behavior of search-engine users who
issue queries related to NASDAQ-100 tickers. In particular, our goal was
to answer to the following questions:

\begin{itemize}
\item What does a typical user search for?
\item Does a user look for many different tickers, or just for a few ones or even one?
\item Does a user ask the same question repeatedly on a certain regular basis, or  sporadically?
\item Can we identify groups of users with a similar behavior?
\end{itemize}

First, we compute the distribution of the number of distinct tickers that any user looks at within a month. 
We then obtain an average monthly distribution by averaging over the 12 months
in our period of observation, as shown in Fig. ~\ref{fig:tickers-per-user}. We also compute the distribution of the
number of distinct tickers that any user looked at within the whole
year, as shown in Fig. ~\ref{fig:tickers-per-user}. The distributions show
very clearly that the overwhelming majority of the users search only
for one ticker, not only within one month, but also within the whole
year.

To further characterize the behavior of users with respect to this one ticker they look for,
we then check how frequently people look for their favorite ticker, and if they search it 
regularly over time (once a day, once a week, once a month).
To conduct this study we focus on three of the tickers characterized by the highest cross-correlation between
query volumes and trading volumes: AAPL (Apple Inc.), AMZN (Amazon.com), and NFLX (NetFlix, Inc.).

For each of these tickers, we consider the set of users who made at least one search related to the ticker during the whole year, 
and we compute the distribution of the number of days on which any users searched the ticker. 
We first consider, separately, the distribution for each month, and then we take the
average over the twelve months. We also compute the distribution over
the whole year. The yearly and monthly distributions for the three tickers are shown in
Figs. ~\ref{fig:AAPL-activity},~\ref{fig:AMZN-activity},~\ref{fig:NFLX-activity}. Surprisingly,
in all the cases considered, a major fraction of the users ($\sim90\%$)
looks at their favorite ticker only one time during a month and the whole year.

Given the correlation and the anticipation of query volumes over
trading volumes described in the previous section one could expect to
observe a significant fraction of users regularly querying for a stock
and doing so more frequently in coincidence of peaks of trading
activity. In contrast, the typical behavior of users suggests the
profile of people who are not financial experts nor regularly
following the market trend. It is thus remarkable that, despite
emerging from the uncoordinated action of ``normal'' people, the query
activity still works well as a proxy to anticipate market trends.

Finally, for the subset of users who have a registered Yahoo! profile, we also analyze the personal data that they provide concerning gender, age, country.
To check if the users who seek NASDAQ-100 tickers behave differently from the rest of the Yahoo! users, we compare the set of registered users who
submitted at least one query related to a NASDAQ-100 ticker with a random sample containing half of the registered users who were tracked in the log 
during the whole year. 
We compute the distributions of the demographic properties for the two aforementioned set of users. 

Table~\ref{tbl:age-random} and~Table \ref{tbl:age-nasdaq} respectively report the age distribution for the random sample and for the set of NASDAQ-100 users.
It is worth to observe that the population of NASDAQ-100 users contains a smaller fraction of old people. Altogether, $72\%$ of the NASDAQ-100 users are people
in working age, while this fraction is equal to $65\%$ in the other sample, which we assume to be a fair representative of the whole set of Yahoo! users.

For what concerns gender, we observe that $55\%$ of the NASDAQ-100 users are males, and $45\%$ are females. The random sample has $52\%$ 
of male users, and $48\%$ of females.  Thus the set of users who searched NASDAQ-100 tickers includes a slightly larger fraction of males. 

For the country distribution, we get similar finding on the two set of users. In both cases, the top-5 states which the users 
come from are California ($13\%$), Texas ($8\%$), New York ($5\%$), Florida ($5\%$) and Illinois ($5\%$).
These fractions are expected, given that the aforementioned states are the most populated within the United States.

\section*{Acknowledgments}
This research was supported by EU Grant FET Open Project 255987 ``FOC''.




\section*{Figure Legends}

\clearpage
\section*{Tables}

\begin{table}[ht]
\caption{
\bf{The 100 traded companies included in the NASDAQ-100 index with their relative ticker.}}
\tiny{
\begin{tabular}{l|l|l}
Activision Blizzard (ATVI) & Adobe Systems Incorporated (ADBE) & Akamai Technologies, Inc (AKAM)\\
Altera Corporation (ALTR) & Amazon.com, Inc. (AMZN) & Amgen Inc. (AMGN)\\
Apollo Group, Inc. (APOL) & Apple Inc. (AAPL) & Applied Materials, Inc. (AMAT)\\
Autodesk, Inc. (ADSK) & Automatic Data Processing, Inc. (ADP)&Baidu.com, Inc. (BIDU) \\
Bed Bath \& Beyond Inc. (BBBY) & Biogen Idec, Inc (BIIB) & BMC Software, Inc. (BMC)\\
Broadcom Corporation (BRCM) & C. H. Robinson Worldwide, Inc. (CHRW)& CA, Inc. (CA) \\
Celgene Corporation (CELG) & Cephalon, Inc. (CEPH) &Cerner Corporation (CERN)\\
Check Point Software Technologies Ltd. (CHKP)& Cisco Systems, Inc. (CSCO)& Citrix Systems, Inc. (CTXS)\\
Cognizant Tech. Solutions Corp. (CTSH) & Comcast Corporation (CMCSA)& Costco Wholesale Corporation (COST)\\
Ctrip.com International, Ltd. (CTRP) & Dell Inc. (DELL) &Dentsplay  International Inc. (XRAY)\\
DirecTV (DTV) & Dollar Tree, Inc. (DLTR) & eBay Inc. (EBAY)\\
Electronic Arts Inc. (ERTS) & Expedia, Inc. (EXPE) & Expeditors Int. of Washington, Inc. (EXPD)\\
Express Scripts, Inc. (ESRX) &F5 Networks, Inc. (FFIV) & Fastenal Company (FAST)\\
First Solar, Inc. (FSLR) &Fiserv, Inc. (FISV)& Flextronics International Ltd. (FLEX)\\
FLIR Systems, Inc. (FLIR) & Garmin Ltd. (GRMN) &Genzyme Corporation (GENZ)\\
Gilead Sciences, Inc. (GILD)& Google Inc. (GOOG)& Henry Schein, Inc. (HSIC)\\
Illumina, Inc. (ILMN) & Infosys Technologies (INFY) &Intel Corporation (INTC)\\
Intuit, Inc. (INTU) & Intuitive Surgical Inc. (ISRG) &Joy Global Inc. (JOYG)\\
KLA Tencor Corporation (KLAC) &Lam Research Corporation (LRCX) &Liberty Media Corp., Int. Series A (LINTA)\\
Life Technologies Corporation (LIFE) &Linear Technology Corporation (LLTC) &Marvell Technology Group, Ltd. (MRVL)\\
Mattel, Inc. (MAT) &Maxim Integrated Products (MXIM) &Microchip Technology Incorporated (MCHP)\\
Micron Technology, Inc. (MU) & Microsoft Corporation (MSFT)& Millicom International Cellular S.A. (MICC)\\
Mylan, Inc. (MYL) &NetApp, Inc. (NTAP)& Netflix, Inc. (NFLX)\\
News Corporation, Ltd. (NWSA) &NII Holdings, Inc. (NIHD)& NVIDIA Corporation (NVDA)\\
OÕReilly Automotive, Inc. (ORLY)& Oracle Corporation (ORCL)& PACCAR Inc. (PCAR)\\
Paychex, Inc. (PAYX) &Priceline.com, Incorporated (PCLN)& Qiagen N.V. (QGEN)\\
QUALCOMM Incorporated (QCOM)& Research in Motion Limited (RIMM) &Ross Stores Inc. (ROST)\\
SanDisk Corporation (SNDK)& Seagate Technology Holdings (STX) &Sears Holdings Corporation (SHLD)\\
Sigma-Aldrich Corporation (SIAL)& Staples Inc. (SPLS)& Starbucks Corporation (SBUX)\\
Stericycle, Inc (SRCL)& Symantec Corporation (SYMC)& Teva Pharmaceutical Industries Ltd. (TEVA)\\
Urban Outfitters, Inc. (URBN)& VeriSign, Inc. (VRSN) &Vertex Pharmaceuticals (VRTX)\\
Virgin Media, Inc. (VMED)& Vodafone Group, plc. (VOD) &Warner Chilcott, Ltd. (WCRX)\\
Whole Foods Market, Inc. (WFMI)& Wynn Resorts Ltd. (WYNN)& Xilinx, Inc. (XLNX)\\
Yahoo! Inc. (YHOO)& & \\
\end{tabular}}
\begin{flushleft}.
\end{flushleft}
\label{table:1}
 \end{table}

\begin{table}[ht]
\caption{
\bf{Average cross-correlation time series for NASDAQ-100 stocks (query: Ticker, volumes: searches).}}
\begin{tabular}{ c ||  c | c | c | c | c | c | c | c | c | c | c  }
$\delta$ &   -5  &  -4  &  -3  &  -2  &  -1  &  0  &  1  &  2  &  3  &  4  &  5 \\
\hline
\hline
CCF &  0.0067&  0.0487&  0.0507&  0.0806&  0.1510&  0.3150&  0.2367&  0.0940&  0.0675&  0.0433&  0.0197\\
\end{tabular}
\label{tbl:nasdaq-ticker}
 \end{table}

\begin{table}[ht]
\caption{
\bf{Average cross-correlation time series for NASDAQ-100 stocks (query: Company name, volumes: searches).}}
\begin{tabular}{ c ||  c | c | c | c | c | c | c | c | c | c | c  }
$\delta$ &   -5  &  -4  &  -3  &  -2  &  -1  &  0  &  1  &  2  &  3  &  4  &  5 \\
\hline
\hline
CCF &  0.0159&  0.0629&  0.0508&  0.0455&  0.0639&  0.1196&  0.1083&  0.0561&  0.0509&  0.0299&  0.0169\\
\end{tabular}
\begin{flushleft} Correlations
are lower than the case in which we consider the queries deriving from the tickers (Table \ref{tbl:nasdaq-ticker}).
\end{flushleft}
\label{tbl:nasdaq-companies}
 \end{table}

\begin{table}[ht]
\caption{
\bf{Average cross-correlation functions for the clean NASDAQ-100  stocks (query: Ticker, volumes: searches) .}}
\begin{tabular}{ c ||  c | c | c | c | c | c | c | c | c | c | c  }
$\delta$ &   -5  &  -4  &  -3  &  -2  &  -1  &  0  &  1  &  2  &  3  &  4  &  5 \\
\hline
\hline
CCF &  0.0176&  0.0604&  0.0657&  0.0993&  0.1816&  0.3641&  0.2700&  0.1145&  0.0834&  0.0540&  0.0312\\
\end{tabular}
\begin{flushleft} By clean 
stocks we mean that we remove those stocks which give rise to spurious queries such as the one containing a common words like 
LIFE or for instance the stock EBAY. In Tables S1 and S2 of Supporting Information we report the cross correlation functions of the 87 stocks on which the average is performed.
\end{flushleft} 
\label{tbl:clean-ticker}
 \end{table}

  \begin{table}[ht]
\caption{
\bf{Average cross-correlation time series for NASDAQ-100 clean stocks (query: Ticker, volumes: users).}}
\begin{tabular}{ c ||  c | c | c | c | c | c | c | c | c | c | c  }
$\delta$ &   -5  &  -4  &  -3  &  -2  &  -1  &  0  &  1  &  2  &  3  &  4  &  5 \\
\hline
\hline
CCF &  0.0078&  0.0344&  0.0501&  0.0736&  0.1482&  0.3194&  0.2349&  0.0876&  0.0623&  0.0345&  0.0151\\
\end{tabular}
\begin{flushleft}  The results from the queries of Yahoo! users 
or from all searches (Table \ref{tbl:clean-ticker}) are almost identical.
\end{flushleft}
\label{tbl:clean-user}
 \end{table}

\begin{table}[ht]
\caption{
\bf{Values of cross-correlation functions for some selected stocks.}}
\begin{tabular}{c  || c l c | c l c | c l c | c l c | c l c | c l c | c l c | c l c | c l c | c l c | c l c }
{\bf Ticker}&{\bf $\delta=$-5}&{\bf $\delta=$-4}  & {\bf $\delta=$-3}  & {\bf $\delta=$-2}  & {\bf $\delta=$-1}  & {\bf $\delta=$0}  & {\bf $\delta=$1}  &{\bf  $\delta=$2}  &{\bf  $\delta=$3}  & {\bf $\delta=$4}  &{\bf $\delta=$5} \\
\hline
\hline
ADBE &  0.08 &  0.12   &  0.14  &  0.19   &  0.47   &  0.83   &  0.51 &  0.19  &  0.09  &  0.10  &  0.11\\
CEPH &  0.16   &  0.26  &  0.22  &  0.14   &  0.32  &  0.80  &  0.44 &  0.24  &  0.12  &  0.13  &  0.15\\
APOL &  0.02  &  0.06   &  0.10 &  0.21  &  0.43  &  0.79  &  0.55 &  0.22 &  0.12  &  0.07  &  0.03\\
NVDA &  0.23  &  0.36 &  0.38   &  0.46   &  0.56 &  0.79 &  0.68 & 0.47   &  0.42   &  0.38 &  0.29\\
CSCO &  0.04  &  0.07   &  0.13  &  0.36  &  0.53  &  0.74  &  0.63  &  0.34   &  0.26  &  0.17  &  0.12\\
AKAM &  -0.04  &  -0.06  &  0.03   &  0.07 &  0.22  &  0.72 &  0.49 &  0.20 &  0.11 &  0.02   &  -0.01\\
NFLX &  0.10   &  0.16   &  0.16 &  0.24  &  0.47   &  0.68  &  0.54 &  0.25  &  0.19 &  0.16  &  0.13\\
ISRG &  0.07   &  0.13 &  0.18   &  0.21  &  0.38  &  0.67   &  0.64 &  0.29  &  0.20  &  0.11   &  0.05\\
RIMM &  0.03  &  0.12   &  0.11  &  0.14   &  0.31 &  0.66  &  0.58 &  0.24   &  0.20  &  0.11  &  0.05\\
FFIV &  0.06  &  0.06 &  0.13   &  0.21   &  0.35  &  0.65   &  0.56 &  0.33   &  0.21  &  0.14  &  0.13
\end{tabular}
\begin{flushleft} The values of the cross-correlation function $r(\delta)$  for $\delta > 0$ is always higher than the value of $r(-\delta)$. From this evidence it appears that query volumes anticipate trading volumes by one or two days. See Tables S1 and S2 of Supporting Information for the complete results for the 87 clean stocks. \end{flushleft}
\label{tbl:best}
 \end{table}

\begin{table}[ht]
\caption{
\bf{Cross-correlation coefficient $r(0)$ between query and trading volumes after removing largest events.}}
\begin{tabular}{ c ||  c | c | c  }
{\bf Ticker} & $r(0)$ & $r(0)-$Top5 & $r(0)-$Top 10\\
\hline
\hline
ADBE & 0.83 & 0.51 & 0.32\\
CEPH & 0.80 & 0.32 & 0.24\\
APOL & 0.79 & 0.55 & 0.46\\
NVDA & 0.79 & 0.70 & 0.64\\
CSCO & 0.74 & 0.56 & 0.46\\
AKAM & 0.72 & 0.51 & 0.39\\
NFLX & 0.68 & 0.62 & 0.62\\
ISRG & 0.67 & 0.57 & 0.55\\
RIMM & 0.66 & 0.59 & 0.52\\
FFIV & 0.65 & 0.55 & 0.50\\
\end{tabular}
\begin{flushleft} We compute the cross-correlation coefficient $r(0)$ between query and trading volumes after removing the days characterized by the highest trading volumes, respectively the top five and top ten events are removed. We note that a significant correlation is still observed for most of the stocks considered. This important test supports the robustness of our findings. See Tables S4 and S5 of Supporting Information for the complete results for the 87 clean stocks.   
\end{flushleft}
\label{tbl:drop} 
 \end{table}
\clearpage

 \begin{table}[!ht]
\caption{
\bf{Average cross-correlation functions between search-engine volumes and signed price returns for the clean NASDAQ-100  stocks (query: Ticker, $\delta = 0$)}}
\begin{center}
\begin{tabular}{ l ||  c | c  }
\textbf{Volume} & \textbf{Price returns} & \textbf{Avg correlation} \\
\hline
\hline
searches & $P_+$ & $0.2650$ \\
searches &  $P_-$ & $-0.2360$ \\
searches &  $P_A$  & $0.2728$ \\
\hline
users &   $P_+$  & $0.2722$ \\
users &  $P_-$ & $-0.1975$ \\
users &  $P_A$  & $0.2446$ \\
\end{tabular}
\end{center}
\label{tbl:signed-price-returns}
 \end{table}

\begin{table}[ht]
\caption{
\bf{Granger causality test.}}
\begin{tabular}{c ||  r | r | r | r | r   }
Dataset & lag (days) & Direction & $ \% p < 5\%$ &  $\% p < 1\%$ & Avg reduction in RSS \\
\hline
\hline
Q (100 tickers) & 1 & Q $\rightarrow$ T & $39\%$ & $29\%$ & $4.37\%$ \\
Q (100 tickers) & 1 & T $\rightarrow$ Q & $15\%$ & $5\%$ & $1.71\%$ \\
U (100 tickers) & 1 & U $\rightarrow$ T & $35\%$ & $25\%$ & $3.55\%$ \\
U (100 tickers) & 1 & T $\rightarrow$ U & $8\%$ & $4\%$ & $1.15\%$ \\
\hline
Q (100 tickers) & 2 & Q $\rightarrow$ T & $52.5\%$ & $40.5\%$ & $7.12\%$ \\
Q (100 tickers) & 2 & T $\rightarrow$ Q & $23.2\%$ & $10.1\%$ & $2.63\%$ \\
U (100 tickers) & 2 & U $\rightarrow$ T & $45.4\%$ & $36.4\%$ & $5.31\%$ \\
U (100 tickers) & 2 & T $\rightarrow$ U & $11\%$ & $6.1\%$ & $2.02\%$ \\
\hline
Q (87 tickers) & 1 & Q $\rightarrow$ T & $45.35\%$ & $33.72\%$ & $4.89\%$ \\
Q (87 tickers) & 1 & T $\rightarrow$ Q & $17.44\%$ & $5.81\%$ & $1.78\%$ \\
U (87 tickers) & 1 & U $\rightarrow$ T & $40.7\%$ & $29.1\%$ & $4\%$ \\
U (87 tickers) & 1 & T $\rightarrow$ U & $9.3\%$ & $4.65\%$ & $1.24\%$ \\
\hline
Q (87 tickers) & 2 & Q $\rightarrow$ T & $57.6\%$ & $41.8\%$ & $7.6\%$ \\
Q (87 tickers) & 2 & T $\rightarrow$ Q & $24.4\%$ & $10.5\%$ & $2.7\%$ \\
U (87 tickers) & 2 & U $\rightarrow$ T & $55.1\%$ & $43.7\%$ & $7.97\%$ \\
U (87 tickers) & 2 & T $\rightarrow$ U & $25.3\%$ & $8.05\%$ & $2.92\%$ \\
\end{tabular}
\begin{flushleft} Adding information about yesterday's query volume reduces the average prediction error (in an autoregressive model) for today's trade volume by about $5\%$,
and for half of the companies the reduction is statistically significant at $1\%$.
\end{flushleft}
 \label{tbl:granger} 
 \end{table}
 
\clearpage

\begin{table}[ht]
\caption{
\bf{Age distribution of users.}}
\begin{tabular}{c | c  }
{\bf Age Range}  & {\bf Fraction of Users}\\
\hline
$< 20$ & $6.8\%$ \\
$20 - 30$ & $22.52\%$ \\
$30 - 40$ & $22.81\%$ \\
$40 - 50$ & $19.87\%$ \\
$> 50$ & $27.90\%$ \\
\end{tabular}
\begin{flushleft} Average age distribution for a random sample collecting half of the data
\end{flushleft}
\label{tbl:age-random}
 \end{table}

\begin{table}[ht]
\caption{
\bf{Age distribution for NASDAQ-100 sample.}}
\begin{tabular}{c | c  }
{\bf Age Range}  & {\bf Fraction of Users}\\
\hline
$< 20$ & $5.2\%$ \\
$20 - 30$ & $26.13\%$ \\
$30 - 40$ & $24.86\%$ \\
$40 - 50$ & $21.02\%$ \\
$> 50$ & $22.78\%$ \\
\end{tabular}
\begin{flushleft} We observe some minor differences between the age of common users 
and the one of the users corresponding to queries belonging to NASDAQ-100 sample. 
\end{flushleft}
\label{tbl:age-nasdaq}
 \end{table}

\clearpage
\section*{Supporting Information}
\setcounter{figure}{0}
\setcounter{table}{0}

\section{Data Analysis and Results: all the NASDAQ-100 stocks}
In this section we report the complete results of the stocks on which the averages shown and discussed in the main paper are performed. 
\begin{figure}[!ht]
\renewcommand{\figurename}{Figure S}
\begin{center}
\includegraphics[scale=0.5]{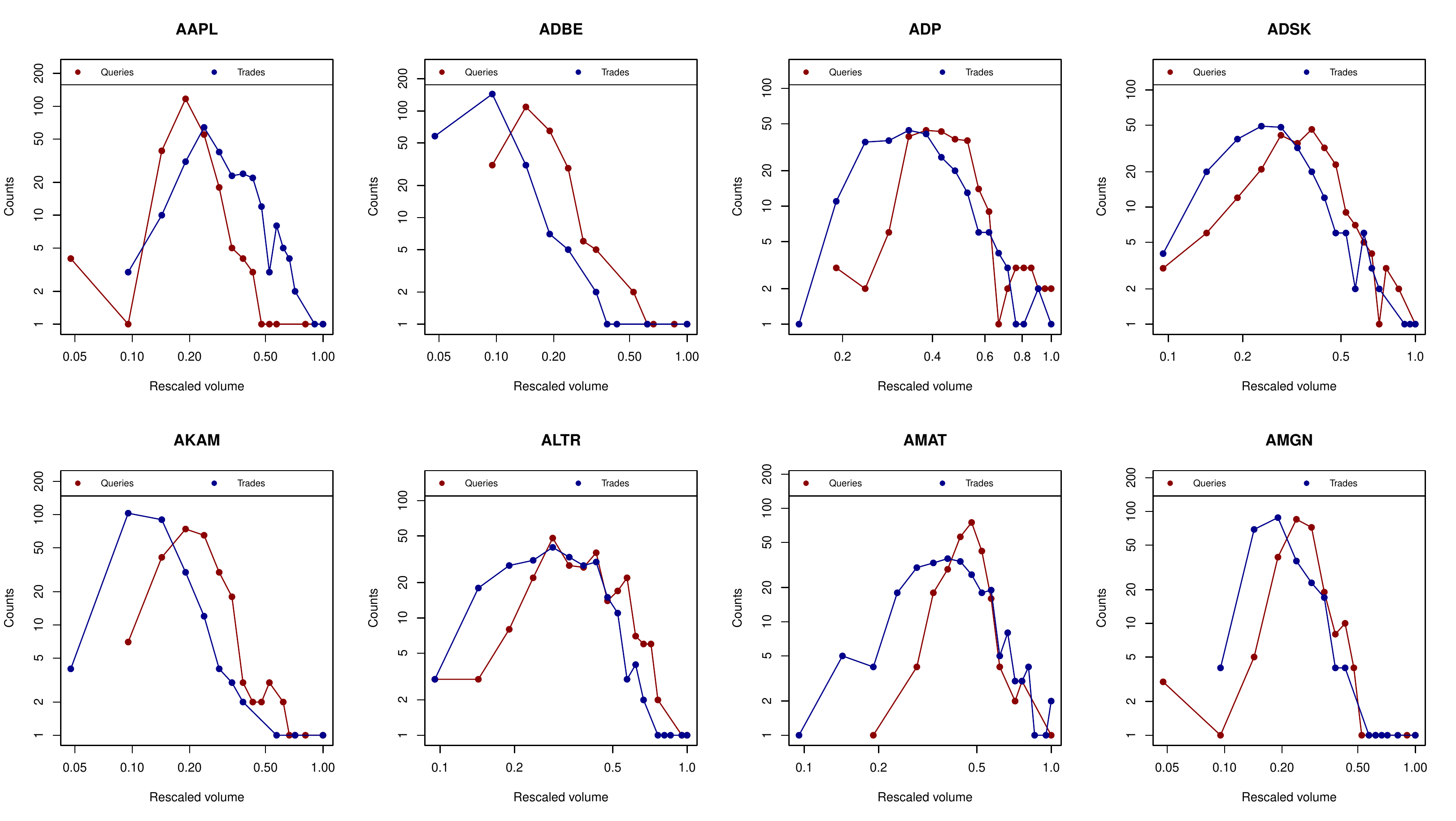}
\includegraphics[scale=0.5]{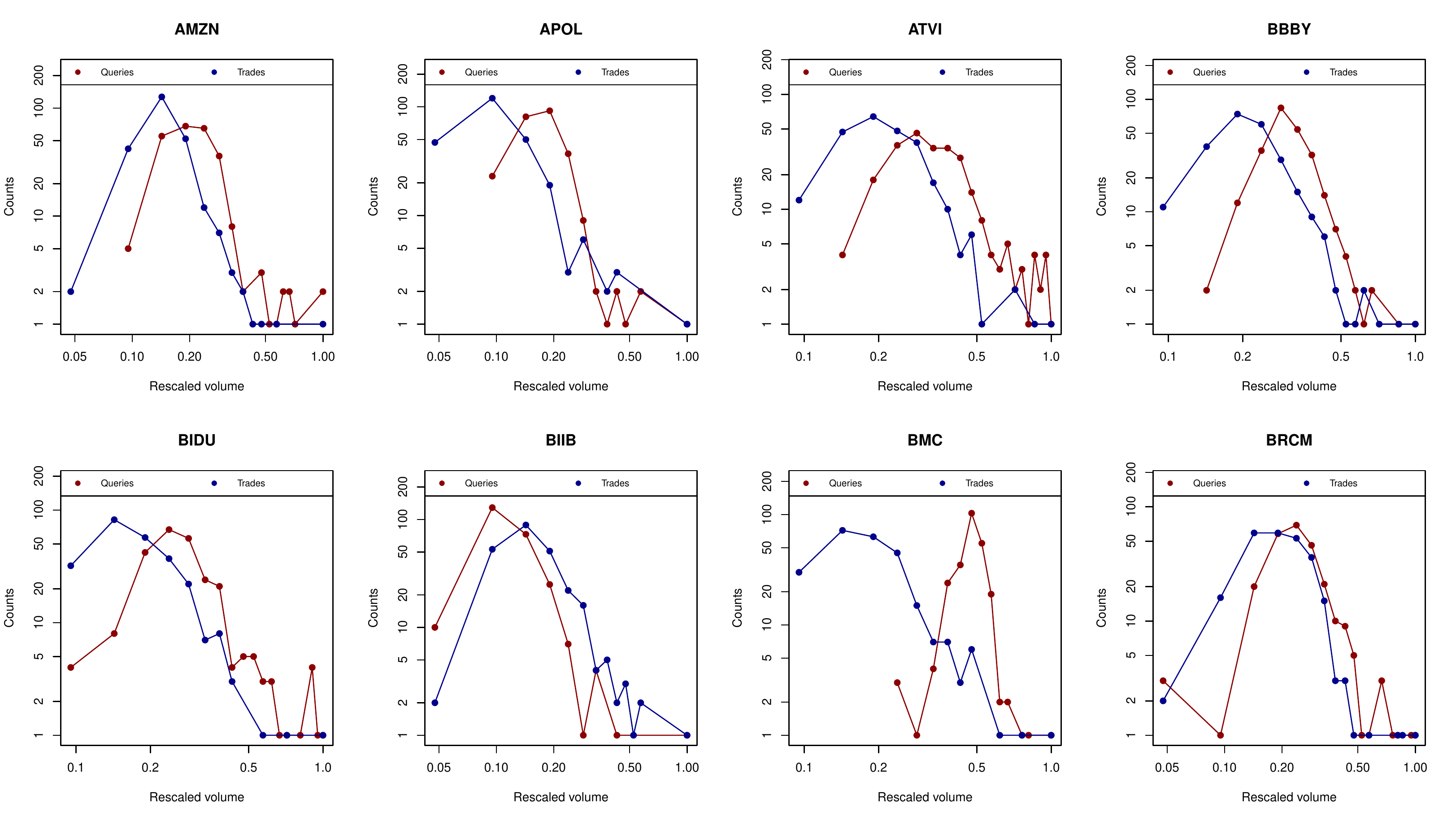}
\end{center}
\caption{\textbf{Histograms of trading volumes and query volumes for all the 87 clean stocks}. Most of the distributions appear to be fat-tailed. All the series have been rescaled dividing them by their maximum value.}
\label{figuretail}
\end{figure}\clearpage

\begin{figure}[!ht]
\renewcommand{\figurename}{Figure S}
\begin{center}
\includegraphics[scale=0.5]{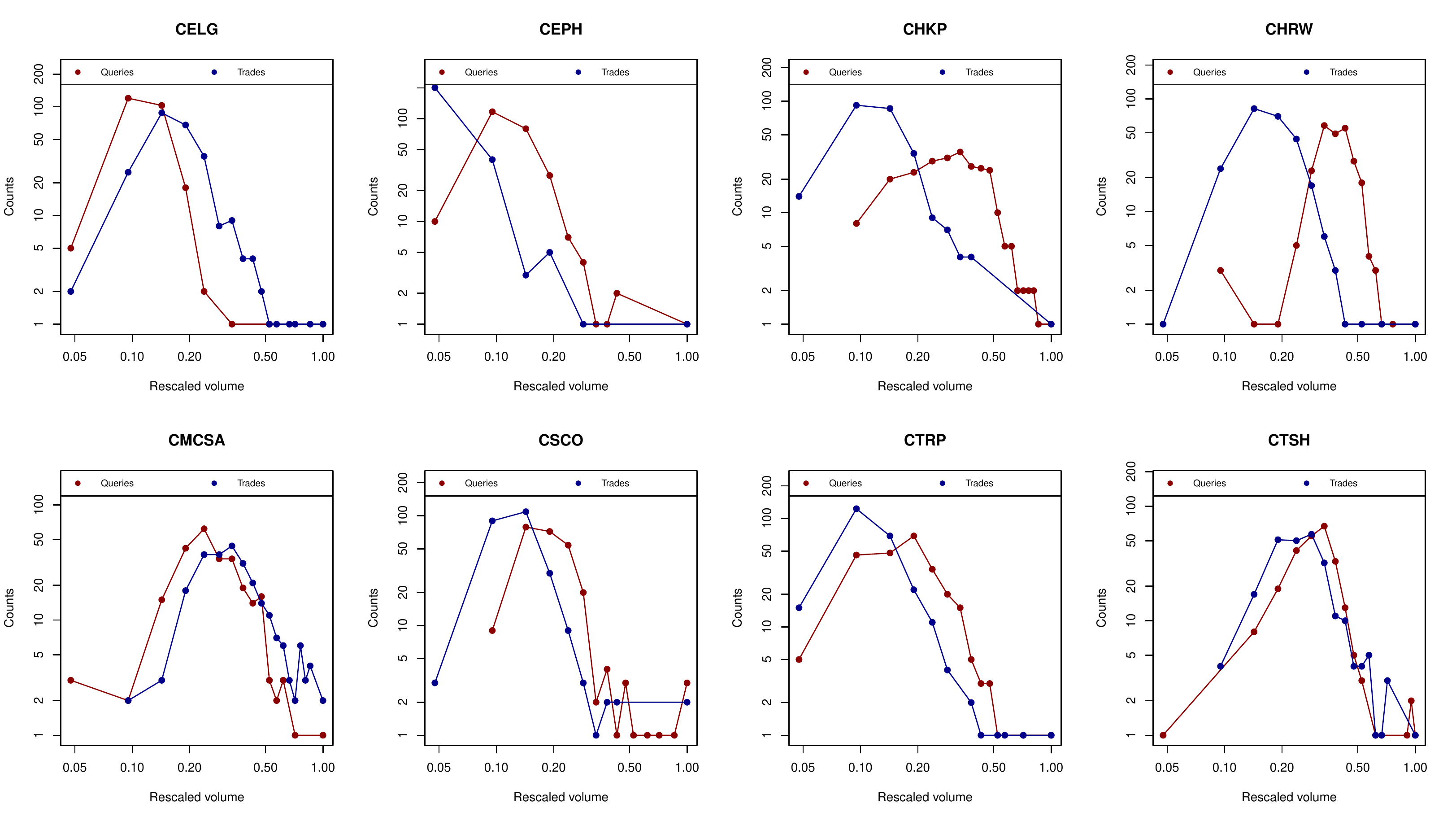}
\includegraphics[scale=0.5]{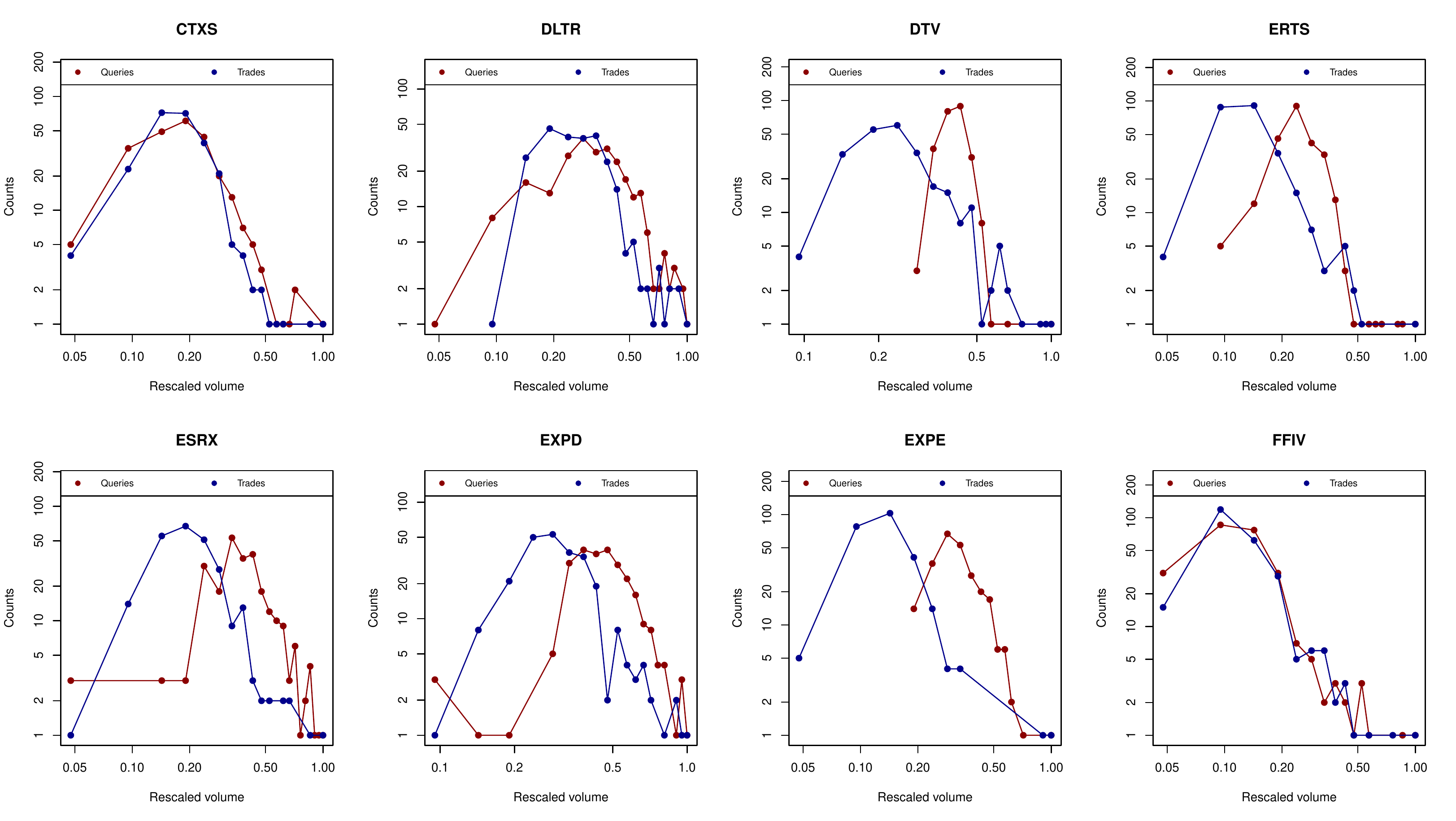}
\end{center}
\caption{\textbf{Histograms of trading volumes and query volumes for all the 87 clean stocks}.  Most of the distributions appear to be fat-tailed.  All the series have been rescaled dividing them by their maximum value.}
\label{figuretail}
\end{figure}\clearpage

\begin{figure}[!ht]
\renewcommand{\figurename}{Figure S}
\begin{center}
\includegraphics[scale=0.5]{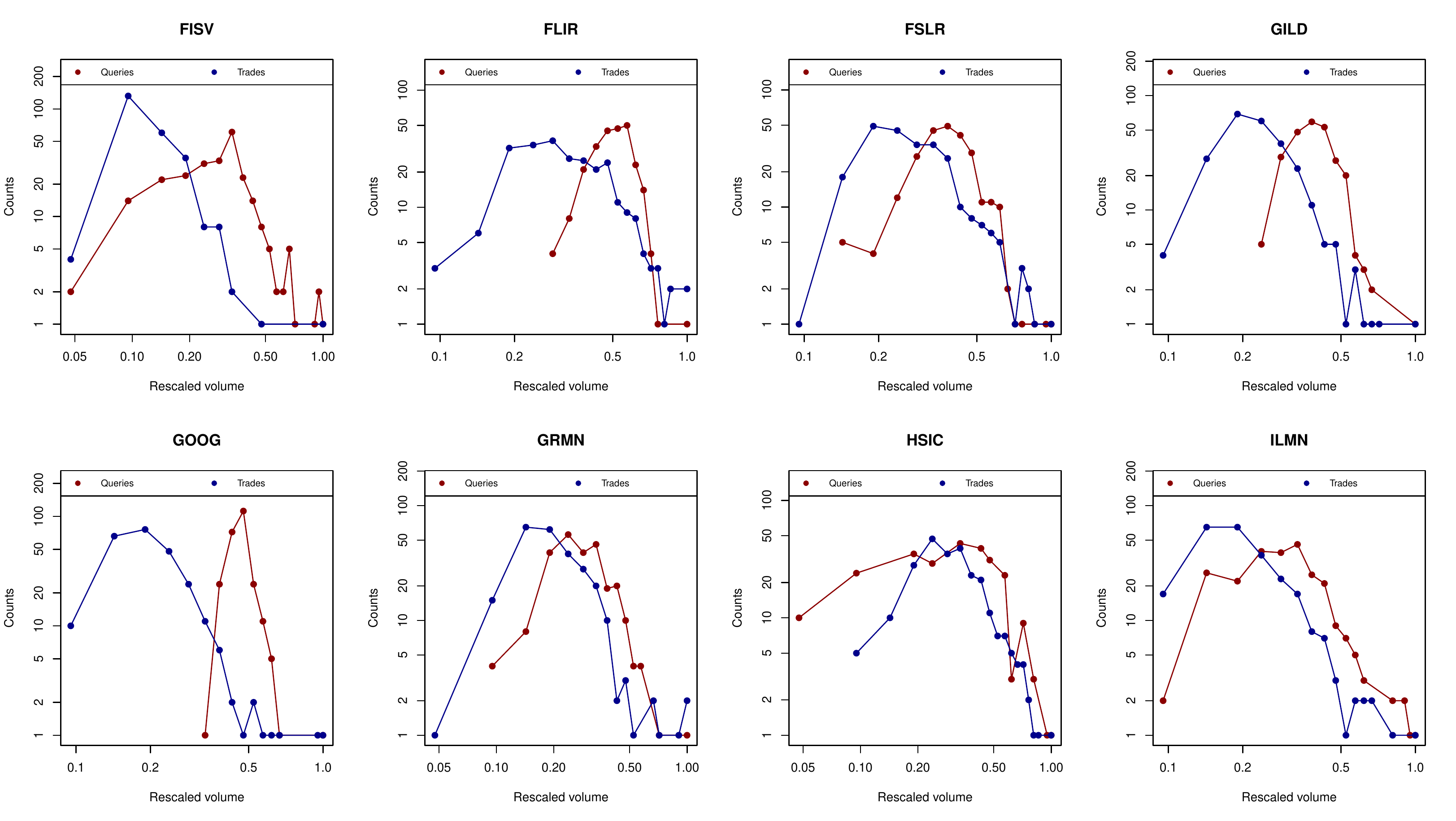}
\includegraphics[scale=0.5]{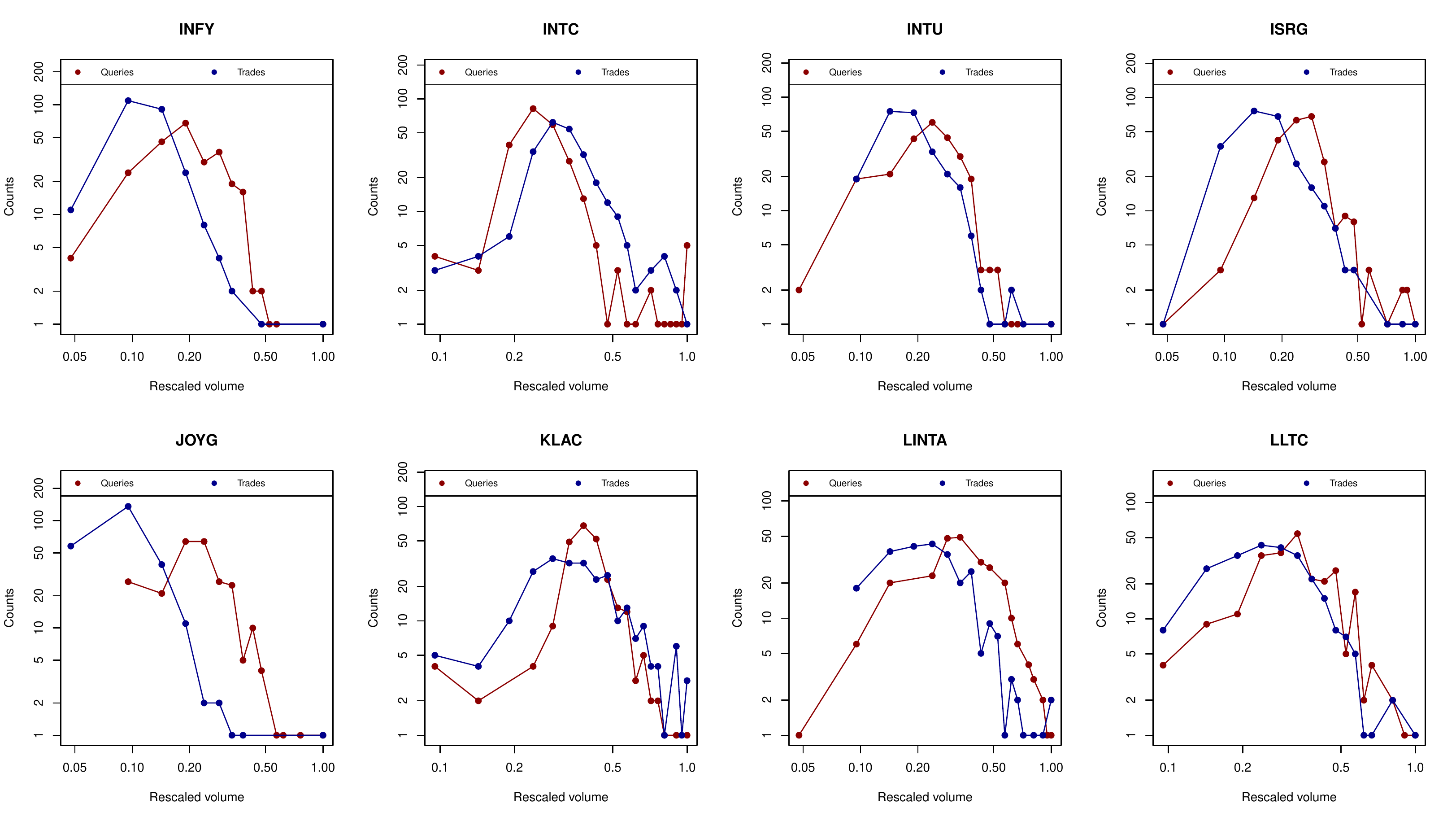}
\end{center}
\caption{\textbf{Histograms of trading volumes and query volumes for all the 87 clean stocks}.  Most of the distributions appear to be fat-tailed. All the series have been rescaled dividing them by their maximum value.}
\label{figuretail}
\end{figure}\clearpage

\begin{figure}[!ht]
\renewcommand{\figurename}{Figure S}
\begin{center}
\includegraphics[scale=0.5]{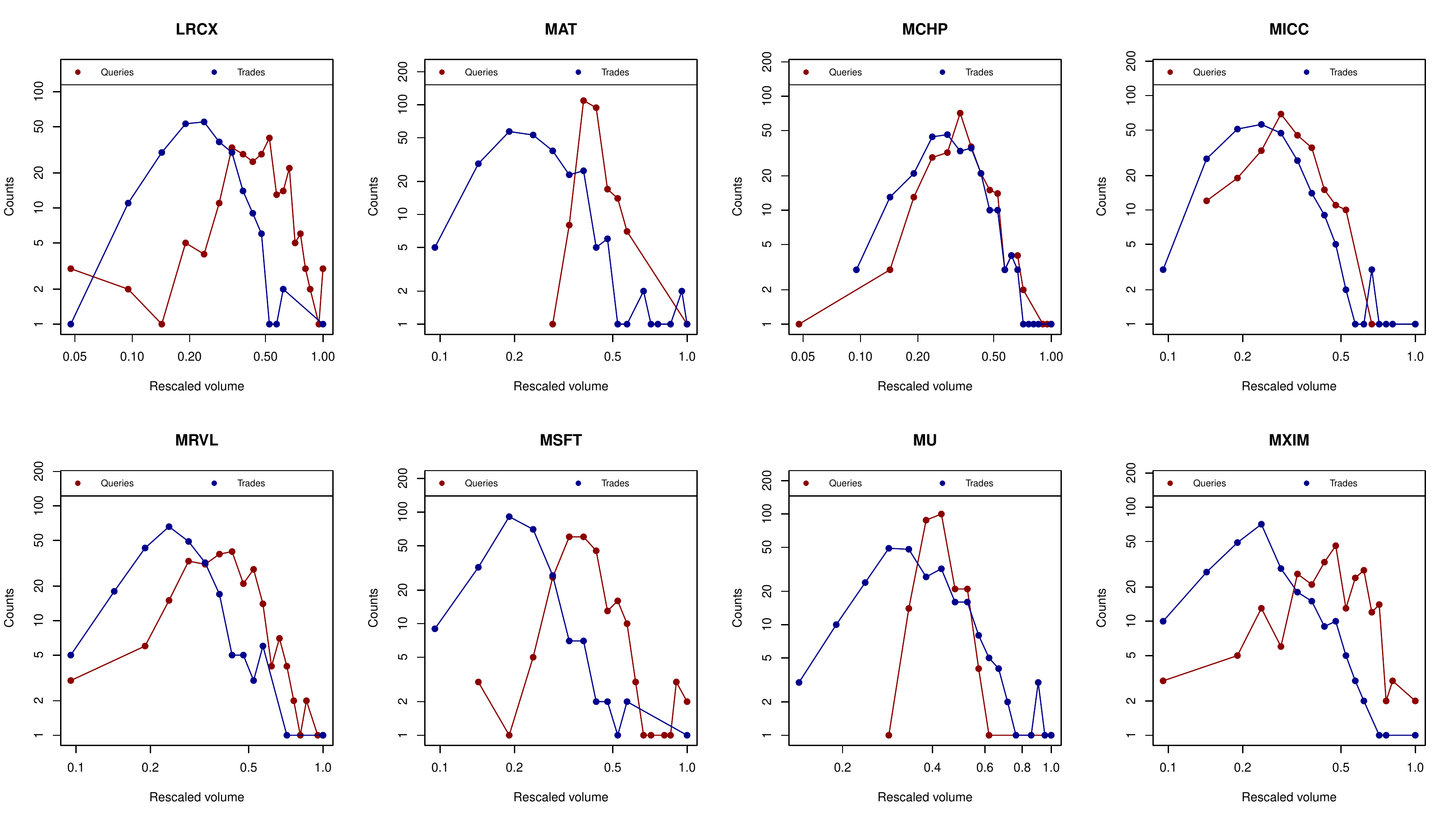}
\includegraphics[scale=0.5]{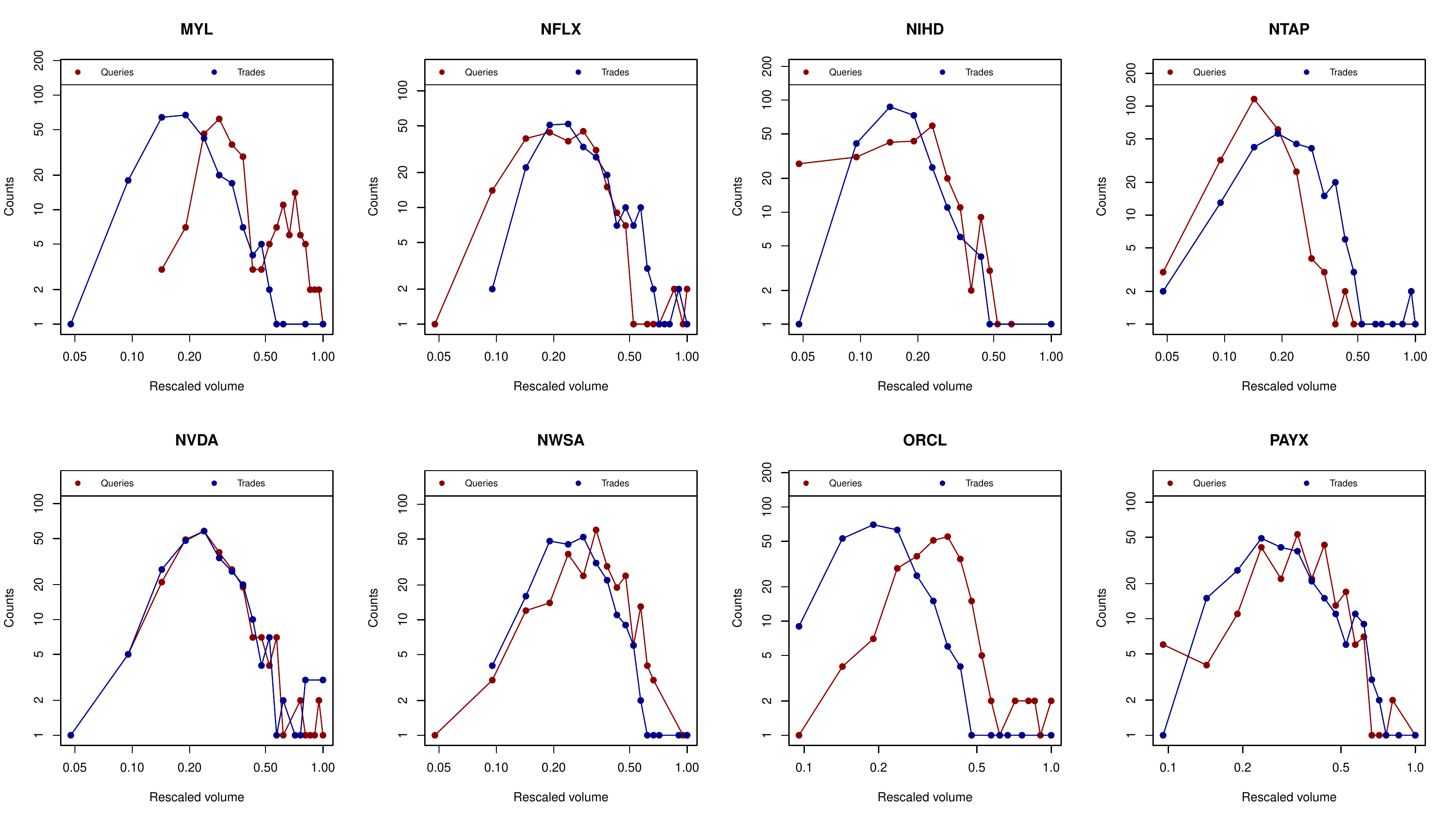}
\end{center}
\caption{\textbf{Histograms of trading volumes and query volumes for all the 87 clean stocks}.  Most of the distributions appear to be fat-tailed. All the series have been rescaled dividing them by their maximum value.}
\label{figuretail}
\end{figure}\clearpage

\begin{figure}[!ht]
\renewcommand{\figurename}{Figure S}
\begin{center}
\includegraphics[scale=0.5]{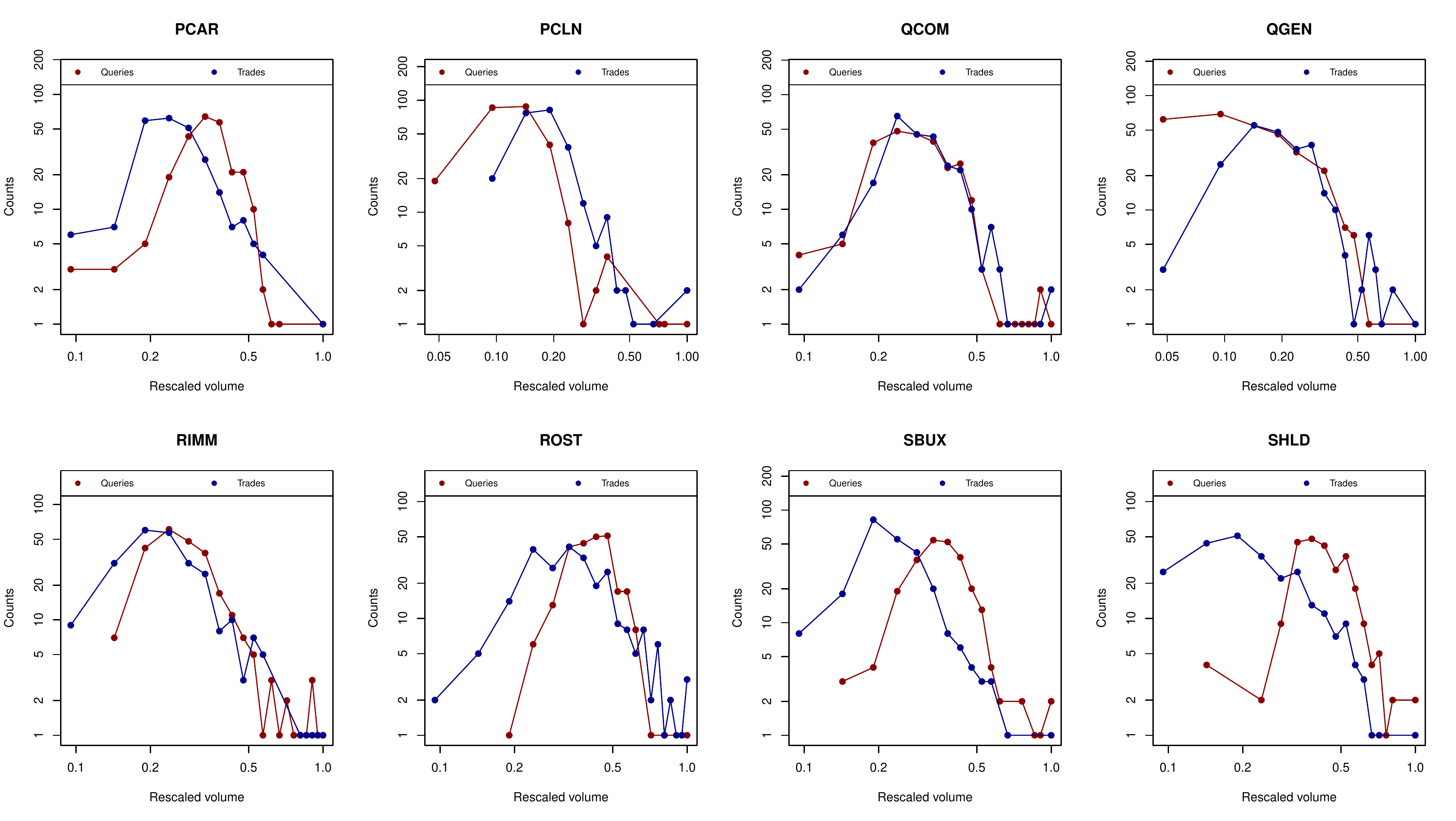}
\includegraphics[scale=0.5]{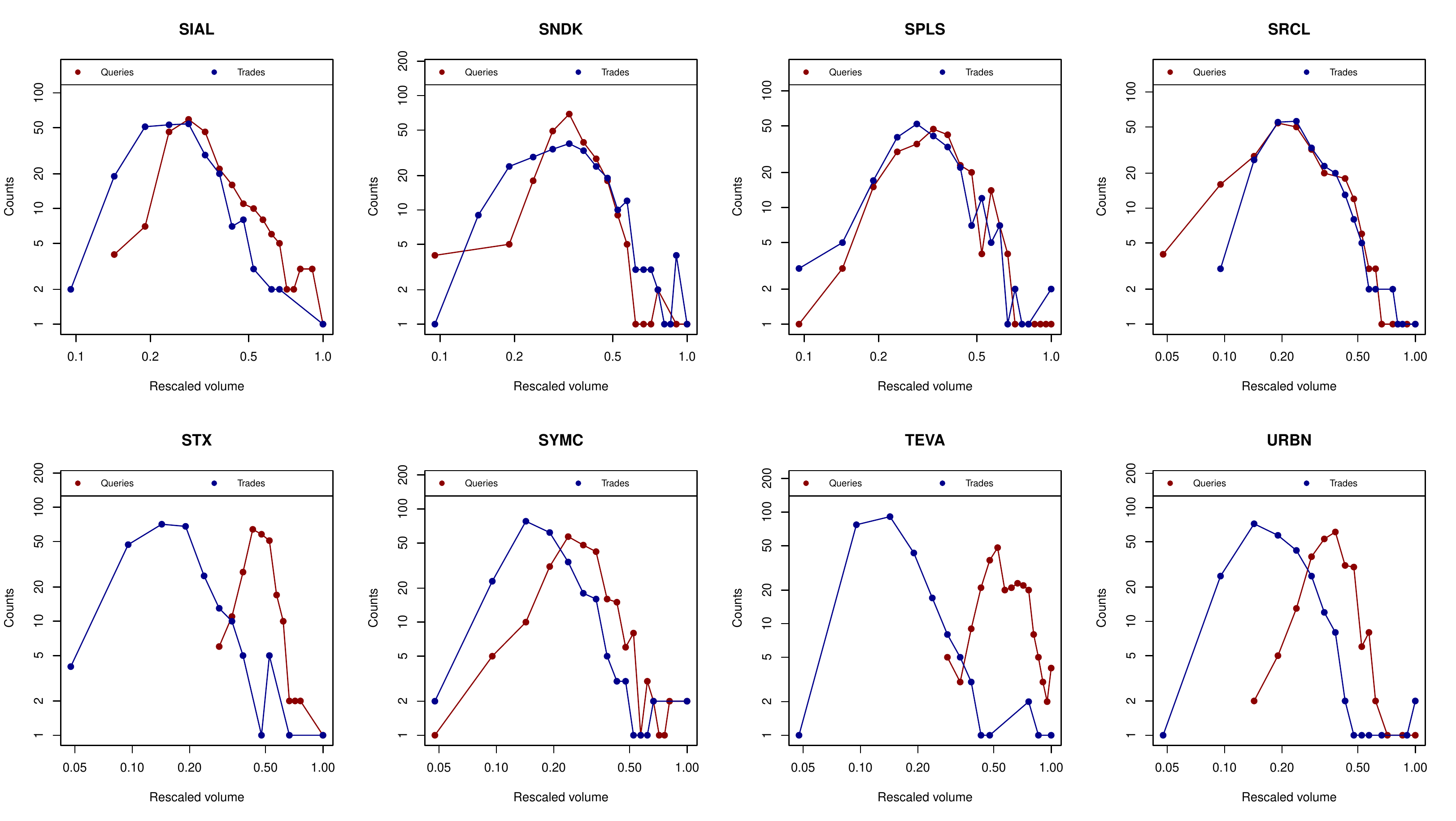}
\end{center}
\caption{\textbf{Histograms of trading volumes and query volumes for all the 87 clean stocks}. Most of the distributions appear to be fat-tailed. All the series have been rescaled dividing them by their maximum value.}
\label{figuretail}
\end{figure}\clearpage

\begin{figure}[!ht]
\renewcommand{\figurename}{Figure S}
\begin{center}
\includegraphics[scale=0.5]{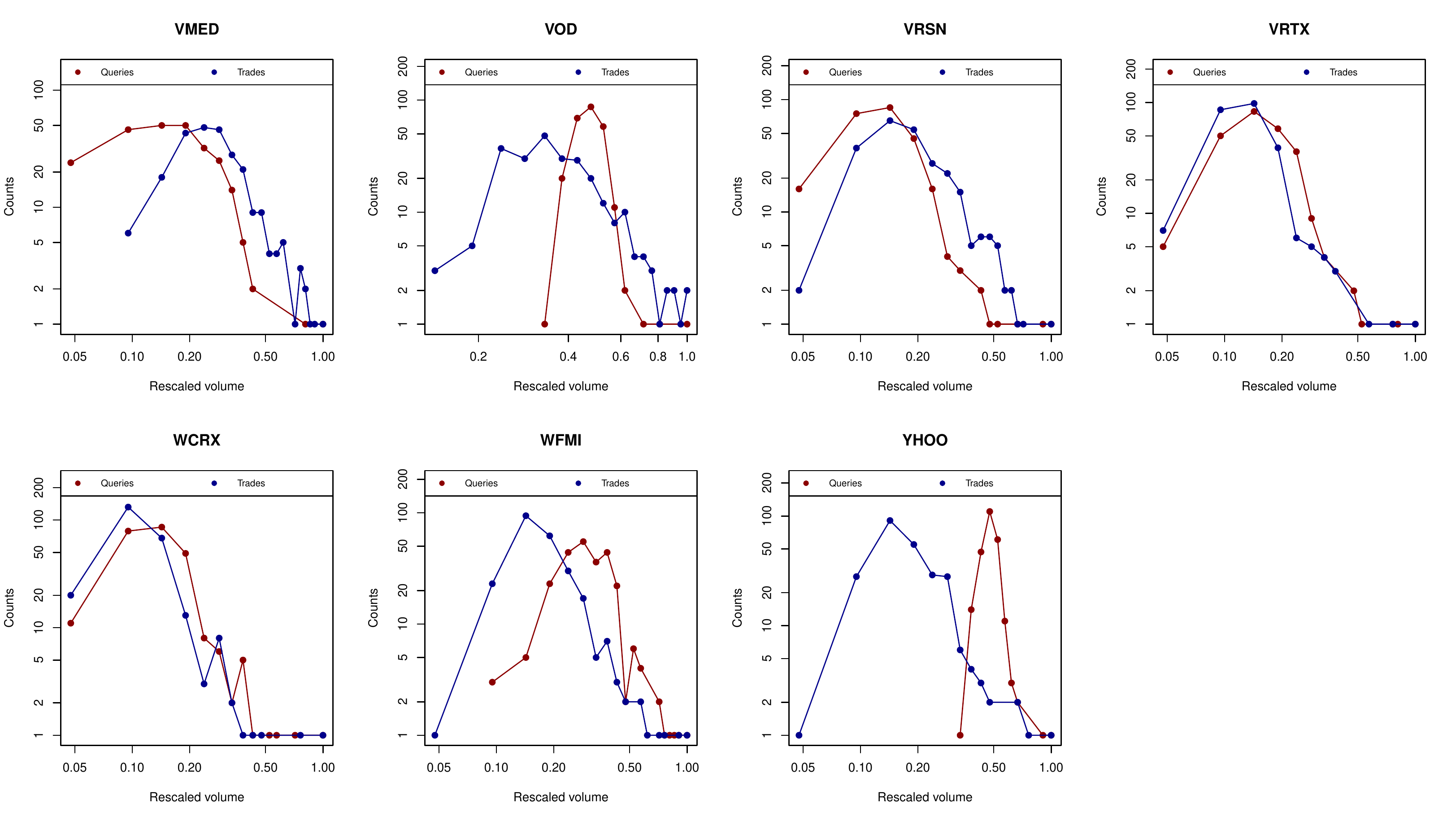}
\end{center}
\caption{\textbf{Histograms of trading volumes and query volumes for all the 87 clean stocks}. Most of the distributions appear to be fat-tailed. All the series have been rescaled dividing them by their maximum value.}
\label{figuretail}
\end{figure}\clearpage

\begin{table}[h!]
\renewcommand{\tablename}{Table S}
\caption{
\bf{Values of cross-correlation functions for the 87 \textit{clean} stocks.}}
\begin{tabular}{c  || c l c | c l c | c l c | c l c | c l c | c l c | c l c | c l c | c l c | c l c | c l c }
{\bf Ticker}&{\bf $\delta=$-5}&{\bf $\delta=$-4}  & {\bf $\delta=$-3}  & {\bf $\delta=$-2}  & {\bf $\delta=$-1}  & {\bf $\delta=$0}  & {\bf $\delta=$1}  &{\bf  $\delta=$2}  &{\bf  $\delta=$3}  & {\bf $\delta=$4}  &{\bf $\delta=$5} \\
\hline
\hline
AAPL  & 0.03  & 0.07  & 0.07  & 0.13  & 0.30  & 0.58  & 0.40  & 0.19  & 0.13  & 0.05  & 0.04\\
ADBE  & 0.08  & 0.12  & 0.14  & 0.19  & 0.47  & 0.83  & 0.51  & 0.19  & 0.09  & 0.10  & 0.11\\
ADP  & -0.19  & -0.21  & -0.23  & -0.16  & -0.15  & -0.15  & -0.18  & -0.19  & -0.15  & -0.14  & -0.15\\
ADSK  & -0.16  & -0.15  & -0.03  & 0.01  & 0.09  & 0.19  & 0.26  & 0.03  & -0.09  & -0.04  & -0.09\\
AKAM  & -0.04  & -0.06  & 0.03  & 0.07  & 0.22  & 0.72  & 0.49  & 0.20  & 0.11  & 0.02  & -0.01\\
ALTR  & 0.34  & 0.40  & 0.40  & 0.37  & 0.42  & 0.55  & 0.53  & 0.39  & 0.41  & 0.37  & 0.38\\
AMAT  & 0.05  & 0.04  & 0.03  & 0.03  & 0.05  & 0.10  & 0.15  & 0.04  & -0.01  & -0.07  & -0.10\\
AMGN  & -0.02  & 0.03  & 0.03  & 0.10  & 0.19  & 0.36  & 0.35  & 0.18  & 0.14  & 0.11  & -0.02\\
AMZN  & -0.07  & -0.03  & -0.04  & -0.02  & 0.13  & 0.48  & 0.43  & 0.04  & -0.02  & 0.01  & -0.02\\
APOL  & 0.02  & 0.06  & 0.10  & 0.21  & 0.43  & 0.79  & 0.55  & 0.22  & 0.12  & 0.07  & 0.03\\
ATVI  & -0.04  & 0.05  & 0.10  & 0.16  & 0.27  & 0.39  & 0.39  & 0.22  & 0.23  & 0.18  & 0.10\\
BBBY  & 0.04  & 0.19  & 0.13  & 0.12  & 0.21  & 0.43  & 0.39  & 0.14  & 0.09  & 0.07  & 0.14\\
BIDU  & 0.10  & 0.09  & 0.12  & 0.19  & 0.32  & 0.49  & 0.42  & 0.16  & 0.11  & 0.05  & 0.04\\
BIIB  & 0.06  & 0.09  & 0.13  & 0.10  & 0.21  & 0.59  & 0.23  & 0.20  & 0.10  & 0.07  & 0.09\\
BMC  & 0.05  & 0.14  & 0.08  & 0.19  & 0.04  & 0.17  & 0.20  & 0.21  & 0.18  & 0.10  & 0.12\\
BRCM  & -0.02  & -0.02  & 0.04  & 0.09  & 0.22  & 0.53  & 0.45  & 0.15  & 0.07  & 0.05  & 0.01\\
CELG  & 0.00  & 0.03  & 0.01  & 0.02  & 0.20  & 0.65  & 0.29  & 0.03  & 0.05  & 0.04  & 0.05\\
CEPH  & 0.16  & 0.26  & 0.22  & 0.14  & 0.32  & 0.80  & 0.44  & 0.24  & 0.12  & 0.13  & 0.15\\
CHKP  & -0.03  & -0.06  & -0.04  & -0.07  & 0.06  & 0.09  & 0.03  & -0.02  & -0.05  & -0.03  & -0.01\\
CHRW  & 0.03  & 0.12  & 0.07  & -0.05  & 0.00  & 0.16  & 0.23  & 0.07  & 0.05  & -0.06  & -0.06\\
CMCSA  & -0.20  & -0.16  & -0.15  & -0.16  & -0.10  & 0.02  & -0.05  & -0.12  & -0.12  & -0.13  & -0.11\\
CSCO  & 0.04  & 0.07  & 0.13  & 0.36  & 0.53  & 0.74  & 0.63  & 0.34  & 0.26  & 0.17  & 0.12\\
CTRP  & -0.02  & 0.08  & 0.01  & 0.11  & 0.19  & 0.57  & 0.26  & 0.06  & 0.03  & 0.04  & 0.06\\
CTSH  & -0.06  & 0.02  & 0.07  & 0.11  & 0.15  & 0.38  & 0.12  & 0.07  & 0.06  & 0.01  & -0.05\\
CTXS  & 0.11  & 0.15  & 0.14  & 0.18  & 0.26  & 0.55  & 0.35  & 0.14  & 0.14  & 0.10  & 0.06\\
DLTR  & 0.07  & 0.16  & 0.17  & 0.14  & 0.23  & 0.42  & 0.25  & 0.24  & 0.15  & 0.07  & 0.04\\
DTV  & 0.04  & 0.02  & 0.06  & 0.09  & 0.05  & 0.03  & 0.03  & -0.05  & 0.05  & 0.03  & -0.01\\
ERTS  & 0.06  & 0.17  & 0.22  & 0.24  & 0.34  & 0.62  & 0.53  & 0.18  & 0.05  & -0.02  & 0.02\\
ESRX  & 0.14  & 0.23  & 0.17  & 0.21  & 0.21  & 0.43  & 0.31  & 0.17  & 0.16  & 0.09  & 0.05\\
EXPD  & 0.01  & 0.07  & 0.08  & 0.09  & 0.24  & 0.37  & 0.31  & 0.22  & 0.18  & 0.16  & 0.11\\
EXPE  & 0.10  & 0.14  & 0.13  & 0.16  & 0.27  & 0.52  & 0.40  & 0.17  & 0.17  & 0.19  & 0.15\\
FFIV  & 0.06  & 0.06  & 0.13  & 0.21  & 0.35  & 0.65  & 0.56  & 0.33  & 0.21  & 0.14  & 0.13\\
FISV  & 0.03  & 0.01  & -0.05  & 0.08  & 0.08  & 0.28  & 0.12  & 0.02  & 0.02  & 0.06  & 0.11\\
FLIR  & -0.13  & -0.11  & -0.10  & -0.15  & -0.14  & -0.13  & -0.09  & -0.12  & -0.16  & -0.16  & -0.16\\
FSLR  & 0.01  & 0.05  & 0.02  & 0.13  & 0.29  & 0.55  & 0.44  & 0.18  & 0.12  & 0.02  & -0.01\\
GILD  & 0.03  & 0.13  & 0.13  & 0.12  & 0.13  & 0.18  & 0.14  & 0.06  & 0.08  & 0.03  & 0.02\\
GOOG  & -0.09  & 0.03  & -0.00  & 0.01  & 0.04  & 0.02  & 0.09  & -0.02  & -0.05  & -0.01  & -0.07\\
GRMN  & 0.07  & 0.10  & 0.07  & 0.05  & 0.23  & 0.46  & 0.24  & 0.12  & 0.09  & 0.07  & 0.08\\
HSIC  & -0.20  & -0.18  & -0.13  & -0.10  & 0.04  & 0.07  & -0.02  & -0.07  & -0.00  & -0.03  & -0.16\\
ILMN  & -0.01  & 0.06  & 0.12  & 0.16  & 0.20  & 0.40  & 0.39  & 0.31  & 0.27  & 0.21  & 0.14\\
INFY  & -0.05  & -0.09  & 0.02  & 0.06  & 0.14  & 0.53  & 0.20  & 0.06  & 0.10  & 0.03  & 0.00\\
INTC  & 0.07  & 0.04  & 0.00  & 0.05  & 0.18  & 0.44  & 0.40  & 0.14  & 0.09  & 0.05  & 0.03\\
INTU  & -0.08  & -0.10  & -0.10  & -0.07  & -0.00  & 0.31  & 0.22  & 0.10  & 0.03  & -0.05  & -0.10\\
ISRG  & 0.07  & 0.13  & 0.18  & 0.21  & 0.38  & 0.67  & 0.64  & 0.29  & 0.20  & 0.11  & 0.05\\
\end{tabular}
\label{tbl:87clean}
 \end{table}
\clearpage

\begin{table}[h!]
\renewcommand{\tablename}{Table S}
\caption{
\bf{Values of cross-correlation functions for the 87 \textit{clean} stocks.}}
\begin{tabular}{c  || c l c | c l c | c l c | c l c | c l c | c l c | c l c | c l c | c l c | c l c | c l c }
{\bf Ticker}&{\bf $\delta=$-5}&{\bf $\delta=$-4}  & {\bf $\delta=$-3}  & {\bf $\delta=$-2}  & {\bf $\delta=$-1}  & {\bf $\delta=$0}  & {\bf $\delta=$1}  &{\bf  $\delta=$2}  &{\bf  $\delta=$3}  & {\bf $\delta=$4}  &{\bf $\delta=$5} \\
\hline
\hline
JOYG  & 0.00  & 0.05  & 0.13  & 0.10  & 0.17  & 0.27  & 0.13  & 0.09  & 0.06  & 0.09  & 0.05\\
KLAC  & 0.36  & 0.40  & 0.40  & 0.46  & 0.45  & 0.43  & 0.49  & 0.46  & 0.47  & 0.43  & 0.39\\
LINTA  & 0.06  & -0.04  & 0.02  & 0.00  & 0.04  & 0.04  & 0.01  & 0.01  & 0.02  & 0.04  & -0.06\\
LLTC  & 0.16  & 0.22  & 0.18  & 0.22  & 0.32  & 0.39  & 0.32  & 0.21  & 0.13  & 0.10  & 0.12\\
LRCX  & -0.01  & 0.00  & -0.02  & 0.04  & 0.17  & 0.24  & 0.20  & 0.16  & 0.14  & -0.03  & 0.00\\
MAT  & 0.06  & 0.21  & 0.07  & 0.10  & 0.09  & 0.04  & 0.06  & 0.03  & -0.02  & 0.05  & 0.02\\
MCHP  & 0.22  & 0.21  & 0.22  & 0.23  & 0.23  & 0.24  & 0.32  & 0.25  & 0.18  & 0.14  & 0.10\\
MICC  & 0.04  & 0.10  & 0.06  & 0.14  & 0.16  & 0.21  & 0.17  & 0.06  & 0.04  & 0.02  & 0.03\\
MRVL  & -0.06  & 0.09  & 0.02  & 0.06  & 0.12  & 0.40  & 0.37  & 0.02  & 0.01  & 0.03  & -0.00\\
MSFT  & -0.09  & -0.02  & -0.06  & 0.02  & 0.17  & 0.42  & 0.35  & 0.02  & 0.05  & 0.04  & 0.09\\
MU  & -0.13  & -0.03  & -0.05  & -0.07  & -0.06  & -0.05  & -0.05  & -0.10  & -0.08  & -0.07  & -0.15\\
MXIM  & 0.11  & 0.09  & 0.19  & 0.18  & 0.22  & 0.29  & 0.11  & -0.04  & 0.03  & 0.03  & 0.01\\
MYL  & -0.10  & -0.07  & -0.07  & -0.10  & -0.11  & -0.07  & -0.07  & -0.06  & -0.07  & -0.04  & -0.01\\
NFLX  & 0.10  & 0.16  & 0.16  & 0.24  & 0.47  & 0.68  & 0.54  & 0.25  & 0.19  & 0.16  & 0.13\\
NIHD  & 0.10  & 0.11  & 0.20  & 0.24  & 0.30  & 0.56  & 0.34  & 0.25  & 0.15  & 0.11  & 0.09\\
NTAP  & -0.06  & 0.02  & 0.01  & 0.06  & 0.26  & 0.61  & 0.46  & 0.18  & 0.09  & 0.09  & 0.11\\
NVDA  & 0.23  & 0.36  & 0.38  & 0.46  & 0.56  & 0.79  & 0.68  & 0.47  & 0.42  & 0.38  & 0.29\\
NWSA  & -0.04  & 0.03  & -0.03  & -0.10  & -0.01  & 0.06  & 0.09  & -0.04  & -0.04  & -0.03  & -0.08\\
ORCL  & 0.09  & 0.17  & 0.09  & 0.07  & 0.23  & 0.52  & 0.43  & 0.13  & 0.16  & 0.10  & 0.03\\
PAYX  & 0.06  & 0.08  & -0.00  & 0.05  & 0.04  & 0.04  & 0.03  & 0.00  & -0.00  & -0.06  & -0.03\\
PCAR  & 0.04  & 0.14  & 0.14  & 0.15  & 0.16  & 0.27  & 0.28  & 0.14  & 0.14  & 0.15  & 0.06\\
PCLN  & -0.10  & -0.04  & -0.03  & 0.01  & 0.20  & 0.51  & 0.37  & 0.06  & -0.01  & -0.06  & -0.06\\
QCOM  & -0.15  & -0.11  & -0.12  & -0.06  & 0.09  & 0.24  & 0.15  & -0.06  & -0.10  & -0.09  & -0.14\\
QGEN  & 0.09  & 0.09  & 0.06  & 0.11  & 0.09  & 0.35  & 0.31  & 0.15  & 0.13  & 0.10  & 0.21\\
RIMM  & 0.03  & 0.12  & 0.11  & 0.14  & 0.31  & 0.66  & 0.58  & 0.24  & 0.20  & 0.11  & 0.05\\
ROST  & -0.22  & -0.12  & -0.15  & -0.11  & -0.17  & -0.08  & -0.12  & -0.10  & -0.13  & -0.20  & -0.16\\
SBUX  & -0.08  & 0.03  & 0.08  & 0.09  & 0.19  & 0.41  & 0.25  & 0.18  & 0.11  & 0.06  & -0.04\\
SHLD  & 0.10  & 0.14  & 0.11  & 0.22  & 0.21  & 0.38  & 0.26  & 0.17  & 0.15  & 0.15  & 0.07\\
SIAL  & -0.05  & 0.00  & -0.02  & -0.03  & -0.05  & -0.05  & 0.00  & -0.01  & 0.01  & -0.01  & 0.02\\
SNDK  & 0.04  & 0.02  & 0.11  & 0.23  & 0.30  & 0.45  & 0.37  & 0.09  & 0.13  & 0.11  & 0.01\\
SPLS  & -0.19  & -0.17  & -0.17  & -0.17  & -0.04  & 0.11  & 0.02  & -0.15  & -0.16  & -0.18  & -0.11\\
SRCL  & 0.05  & 0.02  & -0.04  & 0.01  & 0.12  & 0.27  & 0.24  & 0.21  & 0.08  & 0.05  & 0.05\\
STX  & 0.11  & 0.23  & 0.16  & 0.20  & 0.24  & 0.37  & 0.31  & 0.13  & 0.03  & 0.05  & -0.01\\
SYMC  & -0.00  & 0.02  & 0.11  & 0.17  & 0.25  & 0.58  & 0.44  & 0.21  & 0.14  & 0.04  & 0.04\\
TEVA  & 0.15  & 0.17  & 0.23  & 0.24  & 0.29  & 0.40  & 0.24  & 0.21  & 0.17  & 0.14  & 0.11\\
URBN  & 0.00  & 0.08  & 0.10  & 0.09  & 0.17  & 0.37  & 0.32  & 0.14  & 0.10  & 0.01  & -0.01\\
VMED  & -0.09  & -0.14  & -0.13  & -0.13  & -0.12  & -0.09  & -0.09  & -0.13  & -0.09  & -0.08  & -0.12\\
VOD  & 0.10  & 0.10  & 0.07  & 0.10  & 0.11  & 0.17  & 0.15  & 0.13  & 0.17  & 0.15  & 0.03\\
VRSN  & 0.00  & 0.05  & 0.01  & 0.18  & 0.44  & 0.56  & 0.40  & 0.26  & 0.22  & 0.18  & 0.16\\
VRTX  & 0.02  & 0.14  & 0.32  & 0.42  & 0.30  & 0.50  & 0.24  & 0.07  & 0.19  & 0.14  & 0.16\\
WCRX  & 0.05  & 0.06  & 0.06  & 0.07  & 0.17  & 0.51  & 0.23  & 0.11  & 0.05  & 0.05  & 0.01\\
WFMI  & -0.00  & -0.05  & -0.01  & 0.06  & 0.23  & 0.45  & 0.31  & 0.03  & -0.03  & -0.06  & -0.06\\
YHOO  & 0.06  & 0.15  & 0.15  & 0.16  & 0.23  & 0.38  & 0.25  & -0.02  & -0.00  & 0.02  & -0.03\\
\end{tabular}
\begin{flushleft} The values of the cross-correlation function $r(\delta)$  for $\delta > 0$ are on average larger than the value of $r(-\delta)$. In fact considering only the stocks for which $r(1)>0$ (there are 8 stocks for which $r(1)<0$) we observe that for 68 stocks it holds that $r(1)>r(-1)$ while for the remaining 11 stocks we observe $r(1)\leq r(-1)$.
\end{flushleft}
\label{tbl:87clean2}
 \end{table}
 \clearpage
 
\begin{table}[ht]
\renewcommand{\tablename}{Table S}
\caption{
\bf{Values of cross-correlation functions for the discarded stocks from the original set.} }
\begin{tabular}{c  || c l c | c l c | c l c | c l c | c l c | c l c | c l c | c l c | c l c | c l c | c l c }
{\bf Ticker}&{\bf $\delta=$-5}&{\bf $\delta=$-4}  & {\bf $\delta=$-3}  & {\bf $\delta=$-2}  & {\bf $\delta=$-1}  & {\bf $\delta=$0}  & {\bf $\delta=$1}  &{\bf  $\delta=$2}  &{\bf  $\delta=$3}  & {\bf $\delta=$4}  &{\bf $\delta=$5} \\
\hline
\hline
CERN  & -0.03  & -0.02  & 0.03  & 0.02  & 0.00  & 0.03  & -0.01  & -0.00  & 0.03  & 0.02  & -0.06\\
COST  & -0.28  & -0.21  & -0.18  & -0.19  & -0.20  & -0.17  & -0.06  & -0.07  & -0.16  & -0.10  & -0.11\\
DELL  & -0.05  & -0.04  & -0.01  & 0.01  & -0.11  & -0.11  & -0.05  & -0.04  & -0.05  & -0.02  & -0.07\\
EBAY  & -0.08  & -0.07  & -0.10  & -0.16  & -0.21  & -0.18  & -0.10  & -0.18  & -0.17  & -0.20  & -0.20\\
FAST  & -0.14  & -0.12  & -0.12  & -0.12  & -0.06  & -0.08  & -0.07  & -0.14  & -0.11  & -0.10  & -0.13\\
FLEX  & -0.12  & 0.05  & -0.15  & -0.19  & -0.20  & -0.09  & -0.08  & -0.16  & -0.20  & -0.18  & -0.18\\
LIFE  & -0.07  & 0.01  & -0.04  & -0.01  & -0.09  & -0.08  & -0.05  & -0.06  & 0.01  & 0.11  & 0.11\\
ORLY  & 0.01  & -0.01  & -0.04  & -0.00  & 0.04  & 0.11  & 0.14  & 0.07  & 0.09  & 0.07  & 0.10\\
WYNN  & -0.01  & 0.03  & 0.06  & 0.08  & -0.02  & 0.02  & 0.10  & -0.03  & 0.00  & 0.04  & -0.08\\
XLNX  & 0.03  & -0.00  & 0.02  & 0.06  & 0.09  & 0.14  & 0.14  & -0.03  & -0.03  & -0.12  & -0.03\\
XRAY  & -0.12  & -0.12  & -0.22  & -0.18  & -0.18  & -0.10  & -0.08  & -0.11  & -0.05  & -0.07  & -0.12
\end{tabular}
\begin{flushleft} Most of the query volumes associated to these tickers can be traced back to non-financial origin.  
\end{flushleft}
\label{tbl:noisy}
 \end{table}

\begin{table}[ht]
\renewcommand{\tablename}{Table S}
\caption{
\bf{Cross-correlation coefficient $r(0)$ between query and trading volumes after removing largest events.}}
\begin{center}
\begin{tabular}{ c ||  c | c | c  }
{\bf Ticker} & $r(0)$ & $r(0)-$Top5 & $r(0)-$Top 10\\
\hline
\hline
AAPL & 0.5826 & 0.4769 & 0.4481 \\
ADBE & 0.8326 & 0.5196 & 0.3137 \\
ADP & -0.1456 & -0.1246 & -0.1120 \\
ADSK & 0.1933 & 0.1966 & 0.1795 \\
AKAM & 0.7243 & 0.4893 & 0.4059 \\
ALTR & 0.5546 & 0.5229 & 0.4956 \\
AMAT & 0.1014 & 0.1145 & 0.0732 \\
AMGN & 0.3563 & 0.3165 & 0.3138 \\
AMZN & 0.4838 & 0.3356 & 0.1784 \\
APOL & 0.7927 & 0.5547 & 0.4614 \\
ATVI & 0.3854 & 0.3291 & 0.2410 \\
BBBY & 0.4300 & 0.2963 & 0.2290 \\
BIDU & 0.4891 & 0.3355 & 0.3001 \\
BIIB & 0.5877 & 0.3449 & 0.3320 \\
BMC & 0.1676 & 0.1508 & 0.1600 \\
BRCM & 0.5342 & 0.2219 & 0.2338 \\
CELG & 0.6508 & 0.3171 & 0.1942 \\
CEPH & 0.7959 & 0.3208 & 0.2339 \\
CHKP & 0.0939 & 0.0838 & 0.0808 \\
CHRW & 0.1619 & 0.0559 & 0.0530 \\
CMCSA & 0.0242 & -0.0299 & -0.0456 \\
CSCO & 0.7352 & 0.5614 & 0.5014 \\
CTRP & 0.5659 & 0.3203 & 0.2963 \\
CTSH & 0.3791 & 0.2344 & 0.1756 \\
CTXS & 0.5522 & 0.3525 & 0.2897 \\
DLTR & 0.4243 & 0.3567 & 0.2830 \\
DTV & 0.0308 & 0.0860 & 0.1069 \\
ERTS & 0.6190 & 0.4764 & 0.3225 \\
ESRX & 0.4319 & 0.3371 & 0.2189 \\
EXPD & 0.3749 & 0.3186 & 0.3048 \\
EXPE & 0.5177 & 0.3473 & 0.2712 \\
FFIV & 0.6534 & 0.5410 & 0.5034 \\
FISV & 0.2754 & 0.0568 & 0.0589 \\
FLIR & -0.1267 & -0.1959 & -0.1932 \\
FSLR & 0.5464 & 0.4577 & 0.4020 \\
GILD & 0.1775 & 0.1901 & 0.2013 \\
GOOG & 0.0199 & -0.0440 & -0.1211 \\
GRMN & 0.4564 & 0.2749 & 0.2763 \\
HSIC & 0.0706 & -0.0198 & 0.0053 \\
ILMN & 0.4004 & 0.3020 & 0.3062 \\
INFY & 0.5338 & 0.1080 & 0.0469 \\
INTC & 0.4357 & 0.3178 & 0.3067 \\
INTU & 0.3096 & 0.0262 & -0.0665 \\
ISRG & 0.6683 & 0.5432 & 0.5590 \\
\end{tabular}
\end{center}
\label{tbl:drop-first} 
 \end{table}

\begin{table}[ht]
\renewcommand{\tablename}{Table S}
\caption{
\bf{Cross-correlation coefficient $r(0)$ between query and trading volumes after removing largest events.}}
\begin{center}
\begin{tabular}{ c ||  c | c | c  }
{\bf Ticker} & $r(0)$ & $r(0)-$Top5 & $r(0)-$Top 10\\
\hline
\hline
JOYG & 0.2660 & 0.2147 & 0.1841 \\
KLAC & 0.4307 & 0.4260 & 0.4305 \\
LINTA & 0.0446 & -0.0066 & 0.0156 \\
LLTC & 0.3896 & 0.3286 & 0.2471 \\
LRCX & 0.2424 & 0.2749 & 0.2157 \\
MAT & 0.0441 & -0.0104 & -0.1008 \\
MCHP & 0.2411 & 0.1850 & 0.2042 \\
MICC & 0.2099 & 0.1548 & 0.1556 \\
MRVL & 0.3966 & 0.2554 & 0.2236 \\
MSFT & 0.4216 & 0.3808 & 0.3361 \\
MU & -0.0458 & -0.0440 & -0.0411 \\
MXIM & 0.2948 & 0.2009 & 0.1671 \\
MYL & -0.0665 & -0.0871 & -0.1243 \\
NFLX & 0.6757 & 0.6314 & 0.6253 \\
NHID & 0.5553 & 0.3644 & 0.2925 \\
NTAP & 0.6102 & 0.4173 & 0.2906 \\
NVDA & 0.7856 & 0.6866 & 0.6481 \\
NWSA & 0.0620 & 0.0729 & 0.0794 \\
ORCL & 0.5156 & 0.3493 & 0.3218 \\
PAYX & 0.0365 & 0.1071 & 0.1005 \\
PCAR & 0.2725 & 0.1737 & 0.1798 \\
PCLN & 0.5091 & 0.3054 & 0.2211 \\
QCOM & 0.2444 & 0.0681 & 0.0853 \\
QGEN & 0.3508 & 0.2262 & 0.2092 \\
RIMM & 0.6587 & 0.5946 & 0.5564 \\
ROST & -0.0847 & -0.1247 & -0.1385 \\
SBUX & 0.4095 & 0.3263 & 0.2085 \\
SHLD & 0.3826 & 0.3706 & 0.3563 \\
SIAL & -0.0475 & -0.0053 & -0.0396 \\
SNDK & 0.4510 & 0.3761 & 0.3404 \\
SPLS & 0.1144 & -0.0184 & -0.0031 \\
SRCL & 0.2695 & 0.1365 & 0.1023 \\
STX & 0.3738 & 0.2979 & 0.2242 \\
SYMC & 0.5761 & 0.3703 & 0.4122 \\
TEVA & 0.4005 & 0.2934 & 0.3379 \\
URBN & 0.3714 & 0.2841 & 0.2409 \\
VMED & -0.0938 & -0.1070 & -0.0922 \\
VOD & 0.1682 & 0.1599 & 0.1100 \\
VRSN & 0.5551 & 0.3389 & 0.3199 \\
VRTX & 0.5007 & 0.2135 & 0.1679 \\
WCRX & 0.5106 & 0.3447 & 0.1688 \\
WFMI & 0.4544 & 0.2279 & 0.1042 \\
YHOO & 0.3750 & 0.2145 & 0.1299 \\
\end{tabular}
\end{center}
\begin{flushleft} We compute the cross-correlation coefficient $r(0)$ between query and trading volumes after removing the days characterized by the highest trading volumes, respectively the top five and top ten events are removed. A significant correlation is still observed for most of the stocks considered. \end{flushleft}
\label{tbl:drop-second} 
 \end{table}

\clearpage

\section{Beyond Granger Tests: Tables}
\begin{table}[!ht]
\renewcommand{\tablename}{Table S}
\caption{
\bf{Test 1: p-values for the 26 companies for which $Q \rightarrow V$ at $p=0.01$}}
\begin{center}
\begin{tabular}{ c ||  c | c | c  }
\textbf{Ticker} &	\textbf{$p-val(Q \rightarrow V)$}  & \textbf{$p-val(V \rightarrow Q)$} & \textbf{CCF}\\
\hline
\hline
\textbf{atvi} & $0.000100$ & $0.995100$  &  $0.39$ \\ 
\textbf{csco} & $0.000100$ & $1.000000$  & $0.74$\\
\textbf{expe} & $0.000100$ & $0.997100$  & $0.52$\\
\textbf{ilmn} & $0.000100$ & $0.996800$ & $0.40$ \\
\textbf{isrg} & $0.000100$ & $0.998900$  & $0.67$\\
\textbf{nflx} & $0.000100$ & $1.000000$ & $0.68$\\
\textbf{nvda}  & $0.000100$	& $1.000000$ & $0.79$\\
\textbf{rimm} & $0.000100$	& $1.000000$ & $0.66$\\
\textbf{altr}	& $0.000200$	& $0.999900$ & $0.55$\\
\textbf{msft} & $0.000200$	& $0.961500$ & $0.42$\\
\textbf{symc} & $0.000200$	& $0.986800$ & $0.58$\\
\textbf{mrvl} & $0.000500$ & $0.990800$ & $0.40$\\
\textbf{orcl	} & $0.000500$ & $0.966600$  & $0.52$\\
\textbf{erts} & $0.000800$ & $0.954600$ & $0.62$\\
\textbf{amgn} & $0.000900$ & $0.949500$ & $0.36$\\
\textbf{ffiv}	& $0.001400$	& $0.793400$ & $0.65$\\
\textbf{ntap} & $0.001800$	& $0.968100$  & $0.61$\\
\textbf{bbby} & $0.001900$ & $0.983300$ & $0.43$\\
\textbf{apol} & $0.002000$	& $0.896500$ & $0.79$\\
\textbf{amzn} & $0.002100$	& $0.769600$ & $0.48$\\
\textbf{urbn} & $0.002700$	& $0.880500$ & $0.37$\\
\textbf{vrtx} & $0.003200$	& $0.973900$ & $0.50$\\
\textbf{adbe} &  $0.004300$ & $0.766700$ & $0.85$\\
\textbf{qgen} & $0.006000$ & $0.981300$ & $0.35$\\
\textbf{chrw} & $0.007300$ & $0.965600$ & $0.16$\\
\textbf{stx} & $0.008600$ & 	$0.995600$ & $0.37$\\ 
\end{tabular}
\end{center}
\label{tbl:test1-26}
 \end{table}

\begin{table}[!ht]
\renewcommand{\tablename}{Table S}
\caption{
\bf{Test 1: p-values for the ten companies with smallest $p-val(V \rightarrow Q$ }}
\begin{center}
\begin{tabular}{ c ||  c | c  }
\textbf{Ticker} & $\mathbf{p-val(Q \rightarrow V)}$ & $\mathbf{p-val(V \rightarrow Q)}$ \\
\hline
\hline
\textbf{dltr} & $0.957000$	& $0.019700$ \\
\textbf{mxim} & $0.963700$  & $0.060700$ \\
\textbf{lltc} & $0.936800$ & $0.106900$ \\
\textbf{rost} & $0.851700$	& $0.142600$ \\
\textbf{cmcsa} & $0.913600$ & $0.175900$ \\
\textbf{vrsn} & $0.781100$ & $0.176800$ \\
\textbf{infy} & $0.750700$ & $0.183600$ \\
\textbf{flir}	& $0.836400$ &  $0.193800$ \\
\textbf{vmed} & $0.805600$ & $0.220700$ \\
\textbf{intu} & $0.821200$ & $0.270100$ \\
\end{tabular}
\end{center}
\label{tbl:test1-smallest}
 \end{table}

\begin{table}[!ht]
\renewcommand{\tablename}{Table S}
\caption{
\bf{ List of tickers for which Test 2 gave significant results}}
\begin{center}
\begin{small}
\begin{tabular}{ c ||  c  }
\textbf{Ticker} & \textbf{Outcome} \\
\hline
\hline
aapl & $Q \rightarrow V$ \\
adbe & $Q \rightarrow V$ \\
adp & $Q \rightarrow V$ \\
akam  & $Q \rightarrow V$ \\
altr  & $Q \rightarrow V$ \\
amgn & $Q \rightarrow V$ \\
amzn & $Q \rightarrow V$ \\
apol  & $Q \rightarrow V$ \\
atvi  & $Q \rightarrow V$ \\
bbby  & $Q \rightarrow V$ \\
bidu  & $Q \rightarrow V$ \\
biib & $Q \rightarrow V$ \\
brcm & $Q \rightarrow V$ \\
celg & $Q \rightarrow V$ \\
ceph & $Q \rightarrow V$ \\
csco & $Q \rightarrow V$ \\
ctrp  & $Q \rightarrow V$ \\
ctxs & $Q \rightarrow V$ \\
dltr & $Q \rightarrow V$ \\
erts & $Q \rightarrow V$ \\
esrx & $Q \rightarrow V$ \\
expd & $Q \rightarrow V$ \\
expe  & $Q \rightarrow V$ \\
ffiv  & $Q \rightarrow V$ \\
fslr  & $Q \rightarrow V$ \\
gild  & $Q \rightarrow V$ \\
grmn  & $Q \rightarrow V$ \\
ilmn  & $Q \rightarrow V$ \\
infy  & $Q \rightarrow V$ \\
intc  & $Q \rightarrow V$ \\
iisrg & $Q \rightarrow V$ \\
klac  & $Q \rightarrow V$ \\
lrcx  & $Q \rightarrow V$ \\
mchp  & $Q \rightarrow V$ \\
micc  & $Q \rightarrow V$ \\
mrvl  & $Q \rightarrow V$ \\
msft  & $Q \rightarrow V$ \\
nflx  & $Q \rightarrow V$ \\
nihd  & $Q \rightarrow V$ \\
ntap  & $Q \rightarrow V$ \\
nvda  & $Q \rightarrow V$ \\
orcl  & $Q \rightarrow  V$ \\
pcar  & $Q \rightarrow V$ \\
pcln  & $Q \rightarrow V$ \\
rimm  & $Q \rightarrow V$ \\
sbux  & $Q \rightarrow V$ \\
shld  & $Q \rightarrow V$ \\
sndk  & $Q \rightarrow V$ \\
srcl  & $Q \rightarrow V$ \\
stx  & $Q \rightarrow V$ \\
symc  & $Q \rightarrow V$ \\
urbn  & $Q \rightarrow V$ \\
wcrx  & $Q \rightarrow V$ \\
wfmi  & $Q \rightarrow V$ \\
yhoo  & $Q \rightarrow V$ \\
\end{tabular}
\end{small}
\end{center}
\label{tbl:test-2a}
 \end{table}
 
 \begin{table}[!ht]
 \renewcommand{\tablename}{Table S}
\caption{
\bf{ List of tickers for which Test 2 gave significant results}}
\begin{center}
\begin{small}
\begin{tabular}{ c ||  c  }
\textbf{Ticker} & \textbf{Outcome} \\
\hline
\hline
joyg  & $V \rightarrow Q$ \\
lltc  & $V \rightarrow Q$ \\
rost & $V \rightarrow Q$ \\
teva & $V \rightarrow Q$ \\
vrsn  & $V \rightarrow Q$ \\
vrtx  & $V \rightarrow Q$ \\
\end{tabular}
\end{small}
\end{center}
\label{tbl:test-2b}
 \end{table}

\begin{table}[!ht]
\renewcommand{\tablename}{Table S}
\caption{
\bf{ Outcome of Test 3 }}
\begin{center}
\begin{small}
\begin{tabular}{ c | c | c || c | c | c  }
\textbf{Ticker} & $\mathbf{p-val(Q \rightarrow V)}$ & $\mathbf{p-val(V \rightarrow Q)}$ & \textbf{Ticker} & $\mathbf{p-val(Q \rightarrow V)}$ & $\mathbf{p-val(V \rightarrow Q)}$   \\
\hline
\hline
aapl & $0.000002$ & $0.006796$ & joyg & $0.000000$ & $0.000000$ \\
adbe & $0.000000$ & $0.000000$ & klac & $0.000000$ & $0.000000$ \\
adp &  $0.000000$ & $0.000237$ & linta & $0.030768$ & $0.000000$ \\
adsk &  $0.000000$ & $0.000000$ & lltc & $0.000000$ & $0.000000$ \\
akam &  $0.000000$ & $0.000000$ & lrcx & $0.000000$ & $0.000000$ \\
altr &  $0.000000$ & $0.000000$ & mat & $0.000000$ & $0.000970$ \\
amat &  $0.000000$ & $0.000151$ & mchp & $0.000000$ & $0.000000$ \\
amgn &  $0.000000$ & $0.000545$ & micc & $0.000000$ & $0.000000$ \\
amzn &  $0.000000$ & $0.000000$ & mrvl & $0.000000$ & $0.000000$ \\
apol &  $0.000000$ & $0.000000$ & msft & $0.000000$ & $0.005801$ \\
atvi &  $0.000000$ & $0.000125$ & mu & $0.000002$ & $0.048229$ \\
bbby & $0.000000$ & $0.000061$ & mxim & $0.000000$ & $0.000000$ \\
bidu & $0.000000$ &  $0.000000$ & myl & $0.000000$ & $0.000000$ \\
biib & $0.000000$  & $0.000229$ & nflx & $0.000000$ & $0.000003$ \\
bmc &  $0.000000$ & $0.000000$ & nihd & $0.000000$ & $0.000000$ \\
brcm & $0.000000$  & $0.000000$ & ntap & $0.000000$ & $0.000004$ \\
celg & $0.000575$ & $0.006149$ & nvda & $0.000000$ & $0.000001$ \\
ceph &  $0.000000$ & $0.000000$ & nwsa & $0.000000$ & $0.000000$ \\
chkp & $0.033902$ & $0.000000$ & orcl & $0.000000$ & $0.000000$ \\
chrw & $0.000000$  & $0.000000$ & payx & $0.000027$ & $0.000000$ \\
cmcsa &  $0.000000$ & $0.000000$ & pcar & $0.000000$ & $0.000000$ \\
csco &  $0.000000$ & $0.000000$ & pcln & $0.000000$ & $0.000000$ \\
ctrp &  $0.000000$ & $0.000000$ & qcom & $0.000000$ & $0.000000$ \\
ctsh &  $0.000000$ & $0.000005$ & qgen & $0.000000$ & $0.000000$ \\
ctxs &  $0.000000$ & $0.004119$ & rimm & $0.000000$ & $0.000000$ \\
dltr &  $0.000000$ & $0.000000$ & rost & $0.000000$ & $0.000000$ \\
dtv & $0.000001$ & $0.000001$ & sbux & $0.000000$ & $0.000044$ \\
erts & $0.000000$ & $0.000000$ & shld & $0.000000$ & $0.000000$ \\
esrx & $0.000000$ & $0.000000$ & sial & $0.000080$ & $0.024980$ \\
expd & $0.000000$ & $0.000000$ & sndk & $0.000000$ & $0.000000$ \\
expe & $0.000000$ & $0.000449$ & spls & $0.000006$ & $0.000000$ \\
ffiv & $0.000000$ & $0.000000$ & srcl & $0.000000$ & $0.000000$ \\
fisv & $0.054118$ & $0.000000$ & stx & $0.000000$ & $0.000000$ \\
flir & $0.000005$ & $0.000000$ & symc & $0.000000$ & $0.000001$ \\
fslr & $0.000000$ & $0.000000$ & teva & $0.000000$ & $0.000567$ \\
gild & $0.000000$ & $0.000000$ & urbn & $0.000000$ & $0.000000$ \\
goog & $0.000000$ & $0.155566$ & vmed & $0.000003$ & $0.000000$ \\
grmn & $0.000000$ & $0.000000$ & vod & $0.000000$ & $0.004847$ \\
hsic & $0.000000$ & $0.000000$ & vrsn & $0.000000$ & $0.000000$ \\
ilmn & $0.000000$ & $0.001255$ & vrtx & $0.000000$ & $0.057984$ \\
infy & $0.000000$ & $0.000000$ & wcrx & $0.000000$ & $0.000000$ \\
intc & $0.000000$ & $0.000000$ & wfmi & $0.000000$ & $0.000000$ \\
intu & $0.000000$ & $0.000000$ & yhoo & $0.000000$ & $0.000000$ \\
isrg & $0.000000$ & $0.026642$ & & & \\
\end{tabular}
\end{small}
\end{center}
\label{tbl:test-3}
 \end{table}

\end{document}